\documentclass[fleqn,usenatbib]{mnras}

\usepackage{newtxtext,newtxmath}

\usepackage[T1]{fontenc}
\usepackage{ae,aecompl}

\usepackage{graphicx}	
\usepackage{amsmath}	
\usepackage{amssymb}	

\newcommand{\etal}{et al.\ }
\newcommand{\sigmakpc}{\Sigma_{\text{1 kpc}}}
\newcommand\stmg[1]{$\log$ M$_*/\text{M}_\odot \geq$ #1}
\newcommand\stml[1]{$\log$ M$_*/\text{M}_\odot \leq$ #1}
\newcommand{\sSFR}{$\log$ sSFR}
\newcommand{\som}{$\text{S0}-$\ }

\title[Interdependence of morphology, SFR, and environment]{On the interdependence of galaxy morphology, star formation, and environment in massive galaxies in the nearby Universe}

\author[Bait \etal]
{Omkar Bait,$^{1}$\thanks{E-mail: omkar@ncra.tifr.res.in (OB)} 
    Sudhanshu Barway,$^{2}$\thanks{E-mail: barway@saao.ac.za (SB) }
and Yogesh Wadadekar$^{1}$\thanks{E-mail: yogesh@ncra.tifr.res.in
  (YW)}\\
$^{1}$National Centre for Radio Astrophysics, Tata Institute of Fundamental Research, Post Bag 3, Ganeshkhind, Pune
411007, India \\
$^{2}$South African Astronomical Observatory, P.O. Box 9, 7935, Observatory, Cape Town, South Africa}

\date{Accepted XXX. Received YYY; in original form ZZZ}

\pubyear{2017}

\begin{document}

\pagerange{\pageref{firstpage}--\pageref{lastpage}}
\maketitle

\begin{abstract}
  Using multi-wavelength data, from UV-optical-near-mid IR, for
  $\sim$6000 galaxies in the local Universe, we study the dependence
  of star formation on the morphological T-types for massive galaxies
  ($\log M_*/M_\odot \geq 10$). We find that, early-type spirals (Sa-Sbc) and S0s predominate in the green valley, which is a transition zone between the star forming and quenched regions. Within the early-type spirals, as we move from Sa to Sbc spirals the fraction of green valley and quenched galaxies
  decreases, indicating the important role of the bulge in the
  quenching of galaxies. The fraction of early-type spirals decreases as we enter the green
  valley from the blue cloud, which coincides with the increase in the
  fraction of S0s. This points towards the morphological
  transformation of early-type spiral galaxies into S0s which can
  happen due to environmental effects such as ram-pressure stripping,
  galaxy harassment, or tidal interactions. We also find a second
  population of S0s which are actively star-forming and are present in
  all environments. Since morphological T-type, specific star formation rate (sSFR), and environmental
  density are all correlated with each other, we compute the partial
  correlation coefficient for each pair of parameters while keeping
  the third parameter as a control variable. We find that morphology
  most strongly correlates with sSFR, independent of the environment,
  while the other two correlations (morphology-density and
  sSFR-environment) are weaker. Thus, we conclude that, for massive
  galaxies in the local Universe, the physical processes that shape
  their morphology are also the ones that determine their star-forming
  state.
\end{abstract}


\section{Introduction}

Galaxies are broadly classified, based on their visual morphologies,
into ellipticals (Es), lenticulars (S0s), spirals and irregulars
\citep{Hubble1926}. Remarkably, various physical parameters (in terms
of size, optical colors, luminosity, HI mass fraction, etc.) of
galaxies are known to correlate with morphology
\citep{Roberts1994}. The average star formation rate (SFR) also shows
a strong trend with morphology, with a low rate of star formation in
Es and S0s, and increasing as we go to spiral galaxies
\citep{Kennicutt1998}.

With the advent of large area galaxy surveys, like the Sloan Digital
Sky Survey (SDSS, \citet{York2000}), it was found that in the local
Universe, galaxies show a bi-modal distribution on the optical
color-magnitude diagram \citep{Strateva2001, Kauffmann2003,
  Baldry2004, Baldry2006} with actively star forming ``blue cloud"
galaxies and passively evolving ``red sequence" galaxies. The region
between these two populations is defined as the ``green valley"
\citep{Wyder2007}. It is believed that the green valley of galaxies is
the transition zone between the blue cloud and the red sequence
\citep{Wyder2007,Schiminovich2007,Mendez2011}, and contains galaxies
that have undergone recent quenching of star formation
\citep{Salim2007}. Along with the star-formation properties, green
valley galaxies also have morphological properties (quantified by the
value of the S\' ersic index) intermediate between star-forming and
passive galaxies \citep{Schiminovich2007}. Thus, green valley galaxies
can give us insight into the process of quenching and its dependence
on morphology and other galaxy parameters.

Further, galaxies also show a  correlation between their star
formation rate (SFR) and stellar mass, the so-called `main-sequence'
of star-forming galaxies (MS), both in the local Universe
\citep{Brinchmann2004,Salim2007} and at high redshifts of up to 2
\citep{Daddi2007,Elbaz2007,Noeske2007,Peng2010}. Star-forming galaxies
can loose their position on the MS due to various physical processes
giving rise to a population of passively evolving galaxies. The exact
cause for the cessation of star formation under different physical
conditions is still heavily debated. Phenomenological studies of
quenching have indicated that there are two completely separable
processes : `mass quenching' by which star-forming galaxies undergo
rapid quenching above the Schechter mass ($M^*$), and `environmental
quenching' which produces a second component of quenched galaxies at
the lower mass end \citep{Baldry2006,Peng2010}. Various physical
mechanisms are invoked to explain quenching e.g., halo-quenching
\citep{Dekel2006}, morphological quenching \citep{Martig2009},
AGN-feedback \citep{Croton2006}, strangulation \citep{Larson1980},
ram-pressure stripping \citep{Gunn1972}, and harassment
\citep{Farouki1981,Moore1996}.

Several studies have also shown that the passive state of a galaxy
correlates strongly with the internal structure of a galaxy e.g., with
stellar surface mass density \citep{Kauffmann2003_2}, in particular
with $\sigmakpc$, the inner 1 kpc stellar surface mass density
\citep{Cheung2012, Fang2013, Woo2015} or with bulge mass
\citep{Bluck2014} and also with $\sigma$, the central velocity
dispersion \citep{Wake2012, Teimoorinia2016}. Such a correlation is
more strong for the central galaxy in a dark matter halo than for
satellites. \citet{Omand2014} have shown that quenching depends both on the stellar mass and the effective radius R$_e$. It is important to note that these correlations may not necessarily mean that there is a causal connection between the quenching process and galaxy structural parameters. \citet{Lilly2016} have shown using a simple toy model that such correlations can naturally arise if we consider the \citet{Peng2010} mass-quenching model along with the observed redshift evolution of mass-size relation for star-forming galaxies. Moreover, the passive fraction also shows dependence on
halo mass \citep{Weinmann2006, Wetzel2012, Woo2013, Woo2015}. Further, \citet{Schawinski2014}
found that in the green valley, early type galaxies undergo rapid
quenching whereas the late-type galaxies undergo a slow quenching
process.

Since finding the visual morphologies of millions of galaxies from
large scale galaxy surveys is difficult, several recent studies on
star formation have mostly ignored the morphologies of galaxies. Galaxy Zoo \citep{Lintott2008} has classified the morphologies of a large number of galaxies, but gives a crude classification in terms of only late-type (which includes all types of spiral galaxies) and early-type (ellipticals and S0s) galaxies. On the contrary, in our study we use the detailed visual
morphologies (in 12 types -- ellipticals, S0-, S0, S0/a, Sa, Sab, Sb, Sbc,
Sc, Scd, Sd) from the \cite{Nair2010} catalogue, and hence we can
distinguish between different kinds of spiral galaxies. We can also
differentiate between ellipticals and S0s which the Galaxy Zoo
morphologies do not. 

In this paper, we want to revisit the relation between classical visual
morphologies and star formation in the modern context of the MS of
star-forming galaxies, the transitioning population of green valley
galaxies, and the quenched galaxies, and also as a function of
environment from the low to high densities. We use multiwavelength
data from UV-optical-mid IR and model the SED of $\sim$ 7000 galaxies
to estimate the stellar mass and SFRs. Due to the availability of UV data
we can also define a green valley. In particular, we want to study the
relation between sSFR-morphology, morphology-density, and
density-sSFR. Since all three of the parameters, morphological T-type,
sSFR, and environmental density are known to correlate with each
other, in each of the correlations we also study the differential
effect of the third parameter. Further, in order to determine which
correlation is the strongest, we compute the partial correlation
coefficient for each of the three correlations while keeping the third
parameter as the control variable.

Our paper is organised as follows. In Section \ref{sample and data},
we discuss our sample selection and our multiwavelength data from
UV-optical-IR. We then briefly discuss the spectral energy
distribution (SED) fitting technique using which we derive the stellar
mass and SFR for our whole sample. We also discuss the morphological
classifications and environmental density measurements for our whole
sample. We then show our results and discuss them in Section
\ref{results}. Our main results on the relation between star formation
and morphology are in Section \ref{morphology-sSFR}. In Section
\ref{morphology environment}, we study the morphology-density relation
for our sample of galaxies and separately study this relation for
star-forming, green valley and quenched galaxies. In Section \ref{sSFR
  environment}, we study the relation between the remaining two
parameters, \sSFR\ and environmental density. In Section \ref{partial
  corr coeff}, we then analyse which correlation between the three
parameters, morphological T-type, sSFR, and environmental density is
strongest by computing the partial correlation coefficient. Finally,
we summarize our results and conclude in Section \ref{conclusions}.

Throughout this paper we use the standard concordance
cosmology with $\Omega_M = 0.3$, $\Omega_\Lambda = 0.7$ and $h_{100} =
0.7$.

\section{Sample selection and data analysis} \label{sample and data}

\subsection{Sample selection}

Our sample is drawn from the \citet{Nair2010} (hereafter NA10)
catalogue of detailed visual morphological classification of 14,034
galaxies. NA10 galaxies were selected from the Sloan digital sky
survey (SDSS) DR4 with spectroscopic redshift in the range $0.01 < z <
0.1$, and extinction corrected apparent magnitude limit of $g < 16$
mag. For galaxies in our sample, we use optical imaging data from the SDSS DR12 \citep{Alam2015}
in $u,g,r,i,$ and $z$ bands. In order to have a better constraint on
recent star formation, we make use of Galaxy Evolution Explorer
(GALEX) data in the far-UV (FUV) and near-UV (NUV) filters. We use
near-IR data from the 2 Micron All Sky Survey (2MASS) in $J,H, \text{and }
K$ bands \citep{Skrutskie2006}, wherein we use the model photometry. Mid IR data are taken from the Wide-Field
Infrared Survey Explorer (WISE;\citet{Wright2010}) in all four
channels - W1, W2, W3, and W4.  We construct our sample by cross-matching the NA10 catalogue
with archival data from GALEX, SDSS, 2MASS, and WISE which gives us
7,831 galaxies. Of these, we have flux measurements in all the 14 bands for 6,819 galaxies. For 594 galaxies we have flux measurements in 13 bands due to missing flux in the W4 filter, and for 418 galaxies we have flux measurement in 10 bands due to missing fluxes in all the 4 WISE filters. For all of these galaxies, we also have local environmental density information from \citet{Baldry2006}.

All our sources are resolved in the four WISE bands and are also
associated with the 2MASS extended source catalogue, hence we
calculate the fluxes using elliptical aperture photometry measurement
(w?gmag column in the WISE catalogue).   Furthermore, the WISE pipeline
measurements for resolved sources are known to be systematically
fainter by approximately 0.35 mag, 0.28 mag, 0.44 mag and 0.3 mag in
the W1, W2, W3 and W4 band respectively, and we correct for these
offsets\footnote{\label{note1}\url{http://wise2.ipac.caltech.edu/docs/release/allsky/expsup/sec6_3e.html}} \citep{Brown2014}
before calculating the fluxes. After these corrections are made, as mentioned in the WISE documentation$^{\ref{note1}}$, these fluxes still have an error of $\sim$20\%.

\subsection{Morphological Classification}
\begin{figure}
\begin{center}
\includegraphics[scale=0.55]{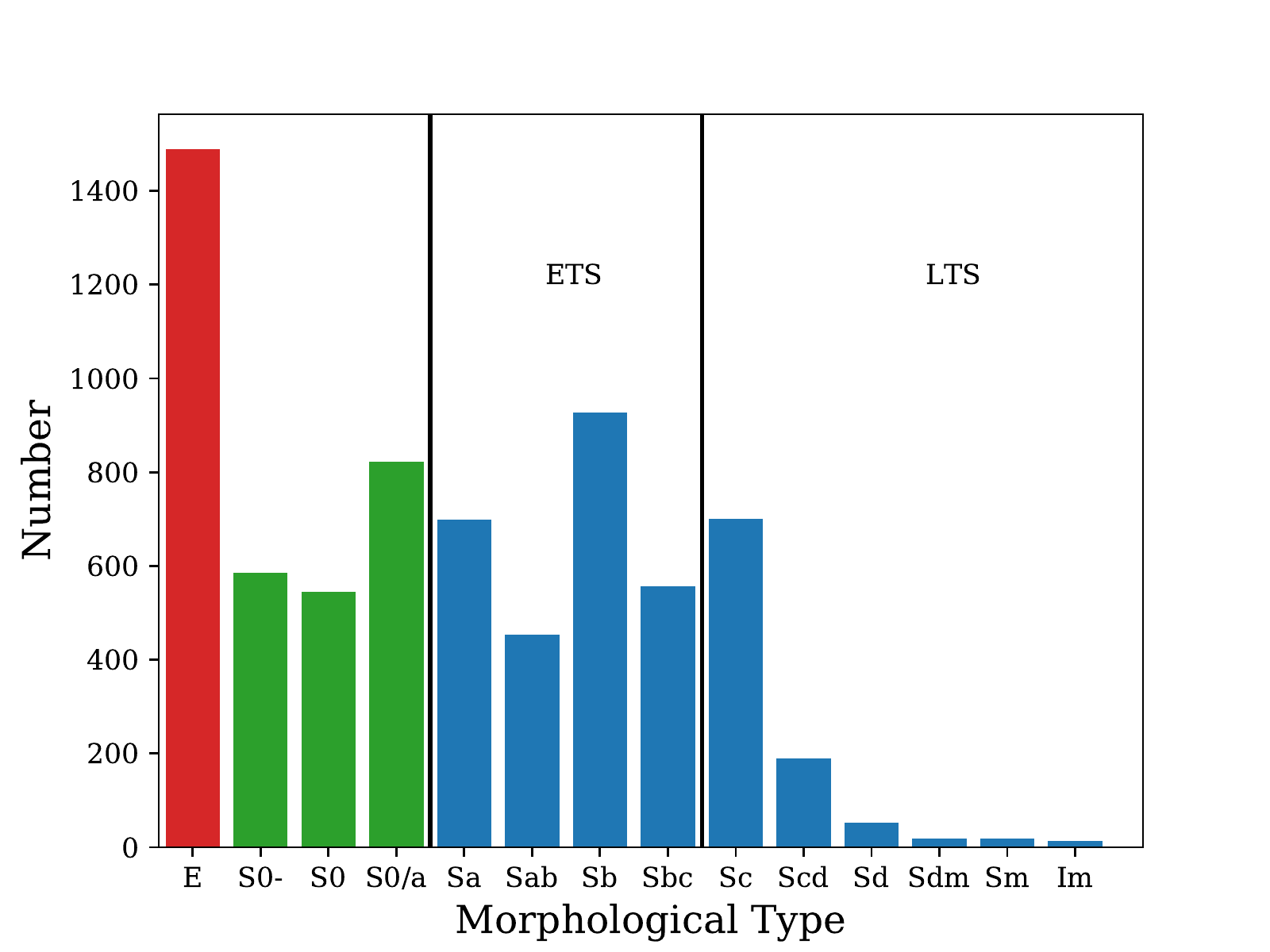}
\caption{Number of galaxies in our sample with various morphologies,  ranging from Es (red), S0s (green), and spirals (blue). We group together spirals ranging from Sa-Sbc and refer to them as early-type spirals (ETS), and Sc-Sd types which we refer as late-type spirals (LTS). Since we have only massive galaxies in our sample, we  have very few number of late-type spiral galaxies. Sdm, Sm, and Im morphologies are removed from our final sample.}\label{morph_hist}
\end{center}
\end{figure}

\begin{table*}
\begin{center}
\caption{Corresponding morphological T-types from the NA10 and RC3 catalogues.}
\begin{tabular}{cccccccccccccccccc}
\hline
Class & c0 & E0 & E+ & S0- & S0 & S0+ & S0/a & Sa & Sab & Sb & Sbc & Sc & Scd & Sd & Sdm & Sm & Im \\
\hline 
NA10 & -5 & -5 & -5 & -3 & -2 & -2 & 0 & 1 & 2 & 3 & 4 & 5 & 6 & 7 & 8 & 9 & 10 \\
\hline
RC3 & -6 & -5 & -4 & -3 & -2 & -1 & 0 & 1 & 2 & 3 & 4 & 5 & 6 & 7 & 8 & 9 & 10 \\
\hline

\end{tabular} \label{T-type mapping}
\end{center}
\end{table*}

The visual morphologies in our final sample are taken from our parent
sample of NA10. In NA10 the authours have done a careful morphological classification for every galaxy using SDSS images in all the five bands u, g,r, i, and z. NA10 can differentiate between different kinds of spirals galaxies into Sa, Sab, Sb, Sbc, Sc, Scd, Sdm, Sm, and Im. It can also differentiate between lenticular galaxies into \som, S0, and S0/a. Galaxies from NA10 which have somewhat doubtful classification (denoted by ?), and galaxies with highly uncertain classification (denoted by :) are not included in our sample. The comparison of the morphological classification
from NA10 with the RC3 catalogue \citep{RC3} is shown in Table \ref{T-type
  mapping}. Since NA10 does not differentiate between c0, E0, and E+
we group them together into ellipticals (Es). Similarly, NA10 does not
differentiate between S0 and S0+ morphologies and hence we group them
together and refer them as S0. Thus, in our classification scheme, we
have three types of lenticular galaxies, S0-, S0, and S0/a. Figure
\ref{morph_hist}, shows the number of galaxies for each of the
morphological classes in our sample. We have very few galaxies
from Sdm-Im morphologies since our sample contains only massive
galaxies and hence we remove them from our sample. This reduces our sample to 7,763 galaxies. The authours in NA10 have classified their entire sample twice and have estimated a mean deviation of less than 0.5 T-types.

We will use such a detailed morphological classification only in
Section \ref{fixed-morph}. For all other sections we will group
together all types of lenticular galaxies (S0-, S0, and S0/a) and
refer them as S0s. And within spiral galaxies we make two groups:
early-type spirals (ETS) ranging from Sa-Sbc types, and late-type
spirals (LTS) ranging from Sc-Sd types.

\subsection{Deriving stellar masses and SFRs using SED fitting}

For each of the 7,763 galaxies in our sample, we model the SED, using
the publicly available Multi-wavelength Analysis of Galaxy Physical
Properties (MAGPHYS) code \citep{daCunha2008} (dC08 hereafter). We
will briefly describe the SED modeling done by MAGPHYS here; for a
detailed description, we refer the reader to dC08.

MAGPHYS creates a library of template spectra and finds the best-fit
spectrum to the available data in UV, optical, near and mid IR
bands. MAGPHYS uses ``CB07 library", which is the unpublished version
of the \citet{Bruzual2003} stellar population synthesis model, with a
Chabrier initial mass function. A set of optical library
templates are constructed, by varying the star formation history,
which is modeled as an exponentially decaying star formation rate with
random bursts of star formation superimposed on it, and also by
varying metallicity from 0.02 to 2 times solar metallicity. The
two-component \citet{Charlot2000} model is used to estimate the amount
of dust attenuation. In this model, the dust attenuation is higher
around young stars (age < $10^7$ Myr) as they are born in dense
molecular clouds. MAGPHYS then performs a $\chi^2$ fit on the given
data, say in UV, optical, and IR bands with the entire library of
template SEDs. For each parameter in the model, MAGPHYS also builds a
likelihood distribution by weighting the parameter value with the
probability given by $\exp(-\chi^2/2)$. We use the median values (the 50th percentile of the marginalized posterior probability distribution function) of stellar mass and SFR for each galaxy in our sample (cf. dc08). For every galaxy we also estimate the uncertainity in estimating the stellar mass and SFR using the 16th and 84th percentile of the marginalized posterior probability distribution, which will be the lower and upper error bars respectively, following dc08. As we lack observations around the 100 micron dust peak,
we are unable to constrain the emission from cold dust, and hence we
refrain from using dust mass and dust luminosity in our analysis.

\begin{figure*}
\centering
\includegraphics[scale=0.9]{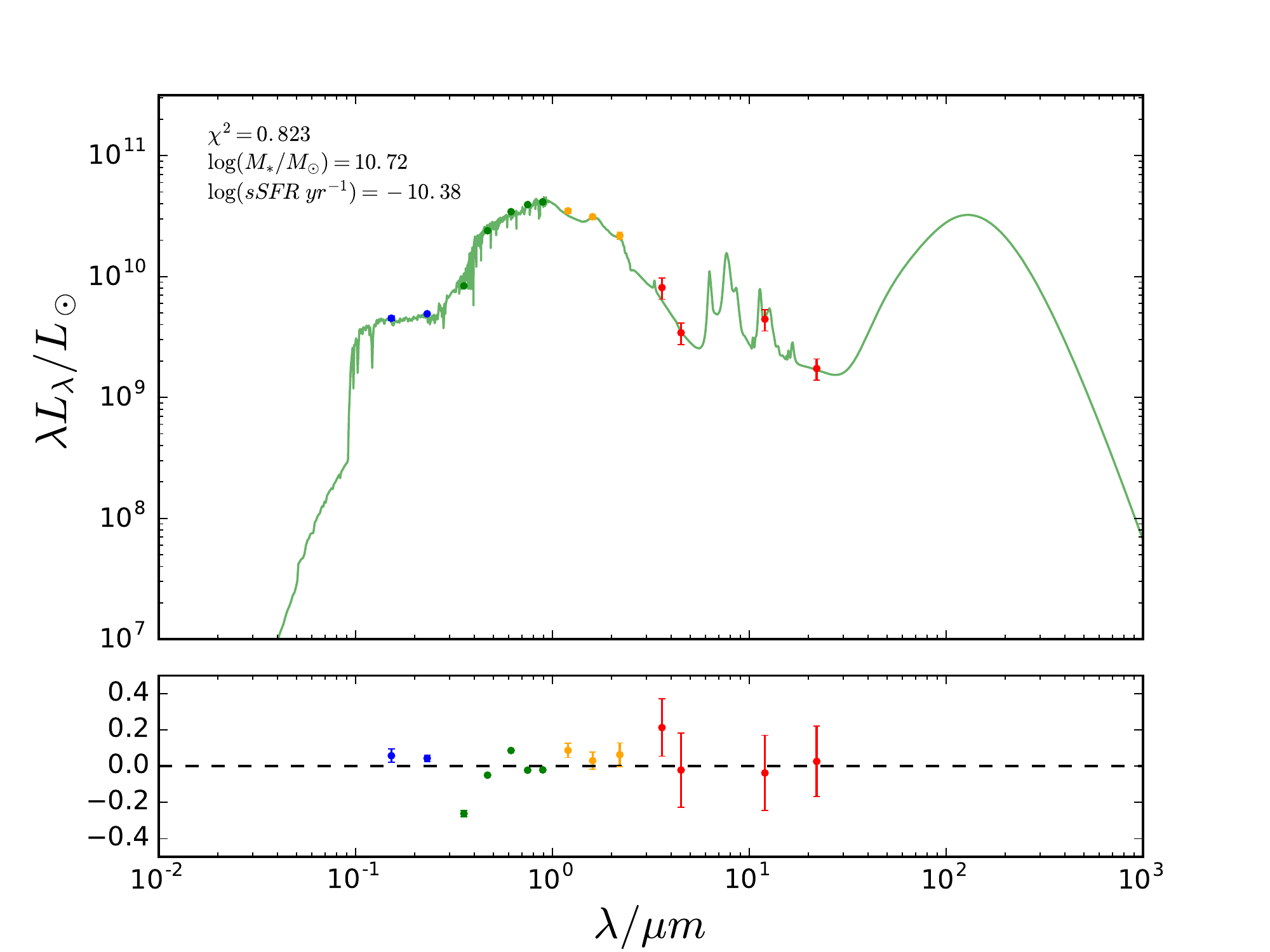}
\caption{Top panel: shows a typical good fit SED in our sample. The x-axis is the rest frame wavelength and y-axis is the luminosity in solar units. The dots are the observed fluxes in 14 bands from GALEX FUV and NUV (blue), SDSS $u,g,r,i,z$ (green), 2MASS J,H,K (yellow) and the four WISE bands (red). GALEX, SDSS, and 2MASS points show 1$\sigma$ errors calculated using the magnitude errors from the corresponding catalogs. The four WISE bands have errors of 20\%. The green line shows the best fit SED. Bottom panel: The dots show the fractional residues $(L_{\text{true}} - L_{\text{MAGPHYS}})/L_{\text{MAGPHYS}}$. The colour code is same as the top panel. Notice that the PAH emission around 10 $\mu$m is not well constrained due to the sparse sampling in this region.  Similarly, the dust emission peak around 100 $\mu$m is not well constrained due to lack of far infrared observations.  }\label{example SED}
\end{figure*}

We used the MAGPHYS SED fitting technique to model the individual SED
using 14 bands, ranging from UV-midIR, of 7,763 galaxies in our
sample.  Following \citet[Eq. B2]{Smith2012}, for a data set of 14
bands, the number of degrees of freedom is 8. This corresponds to a
critical $\chi^2$ of 20, above which there is less than 1 percent
probability that our observations are consistent with the
models. However, following the suggestion of \citet{Hayward2015} of
using a more conservative value than that derived from
\citet{Smith2012}, we use a critical $\chi^2$ of 5, and remove all
galaxies with higher $\chi^2$ from our sample. This criterion reduces the  sample to 7020 galaxies. Figure \ref{example SED} is a typical example of a good fit SED
in our sample. Notice that due to lack of data points the polycyclic
aromatic hydrocarbon (PAH) emission features and FIR dust emission are
not very well constrained.

\begin{figure}
\begin{center}
\includegraphics[scale=0.45]{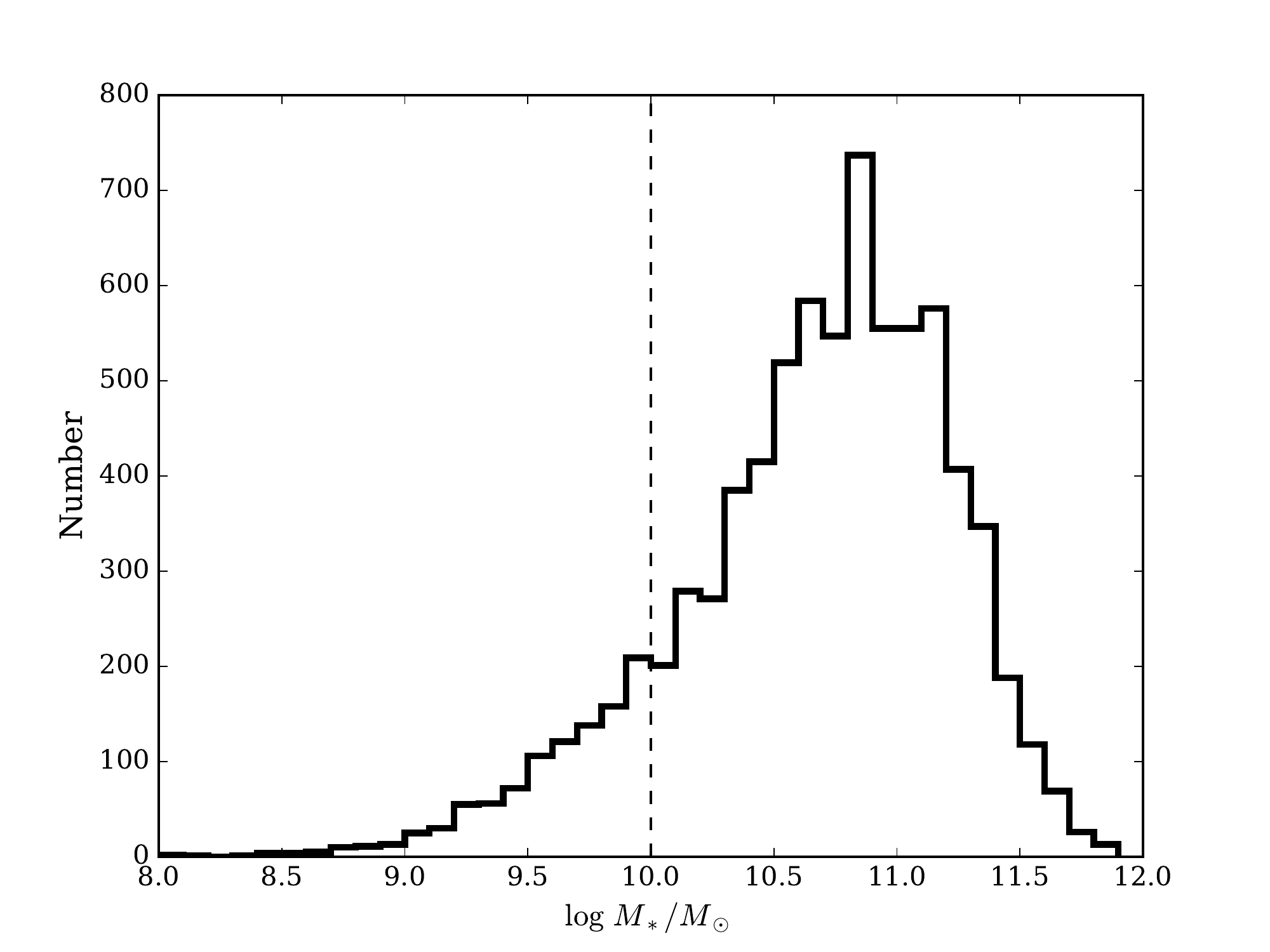}
\caption{Histogram of stellar mass derived using SED fitting from MAGPHYS. In the final  sample only galaxies with \stmg{10}) are included.}\label{M_star_hist}
\end{center}
\end{figure}

\begin{figure}
\begin{center}
\includegraphics[scale=0.45]{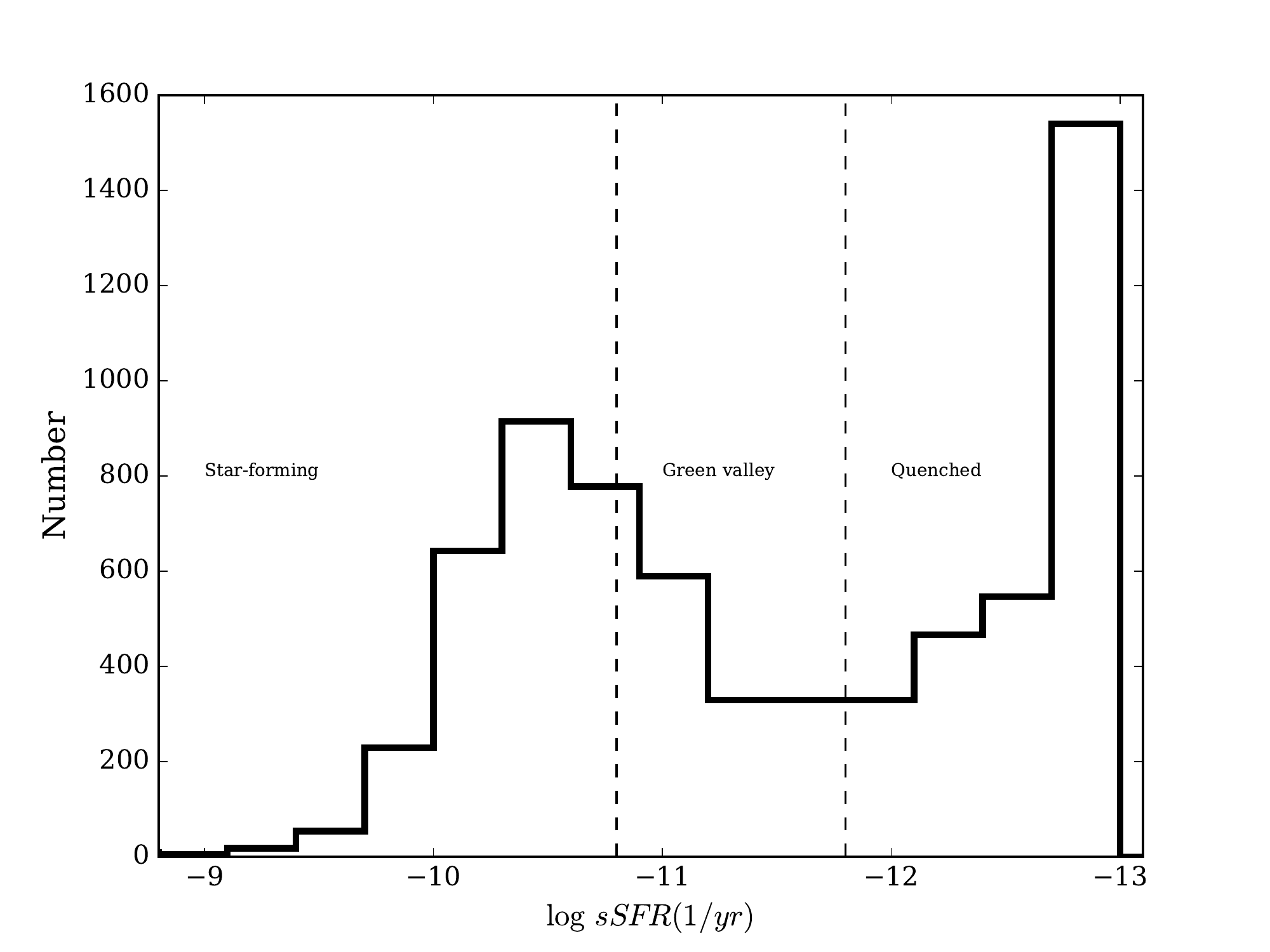}
\caption{Histogram of sSFR, obtained by dividing the SFR with the stellar mass for each galaxy (note that the x-axis is inverted), showing a bi-modal distribution. We define the star-forming region (\sSFR\ $\geq -10.8$), green valley (-10.8 < \sSFR\ < -11.8), and quenched region (\sSFR\ $\leq -11.8$) following \citet{Salim2014}.}\label{sSFR_hist}
\end{center}
\end{figure}

Figure \ref{M_star_hist}, shows the histogram of stellar mass for the sample for 7020 galaxies. As our focus in this paper is only on massive galaxies and their dependence on sSFR, morphology and environment. Hence, we restrict our sample to only massive galaxies (\stmg{10}). The dashed vertical line in Figure \ref{M_star_hist} shows the stellar mass cut. This finally reduces our sample to 6194 galaxies, which we will use in this work.  We caution the reader that the results presented in this paper are valid only for massive galaxies; these could significantly change for lower mass galaxies, where different physical processes may play a role in shaping their  nature. Figure \ref{sSFR_hist}, shows the
distribution of sSFR for our final sample. The sSFR is obtained by
dividing the SFR by the stellar mass for each galaxy. The sSFR is a
proxy for the star formation history of the galaxy, where a high value
of sSFR suggests recent star formation and a low value is an indicator
of an older stellar population with little or no recent star
formation. Therefore, following \citet{Salim2014}, we define star-forming
galaxies with \sSFR\ $\geq -10.8$, green valley galaxies with -11.8 <
\sSFR\ < -10.8, and quenched galaxies with \sSFR\ $\leq -11.8$. 

\subsection{Environmental Density}
\begin{figure}
\begin{center}
\includegraphics[scale=0.45]{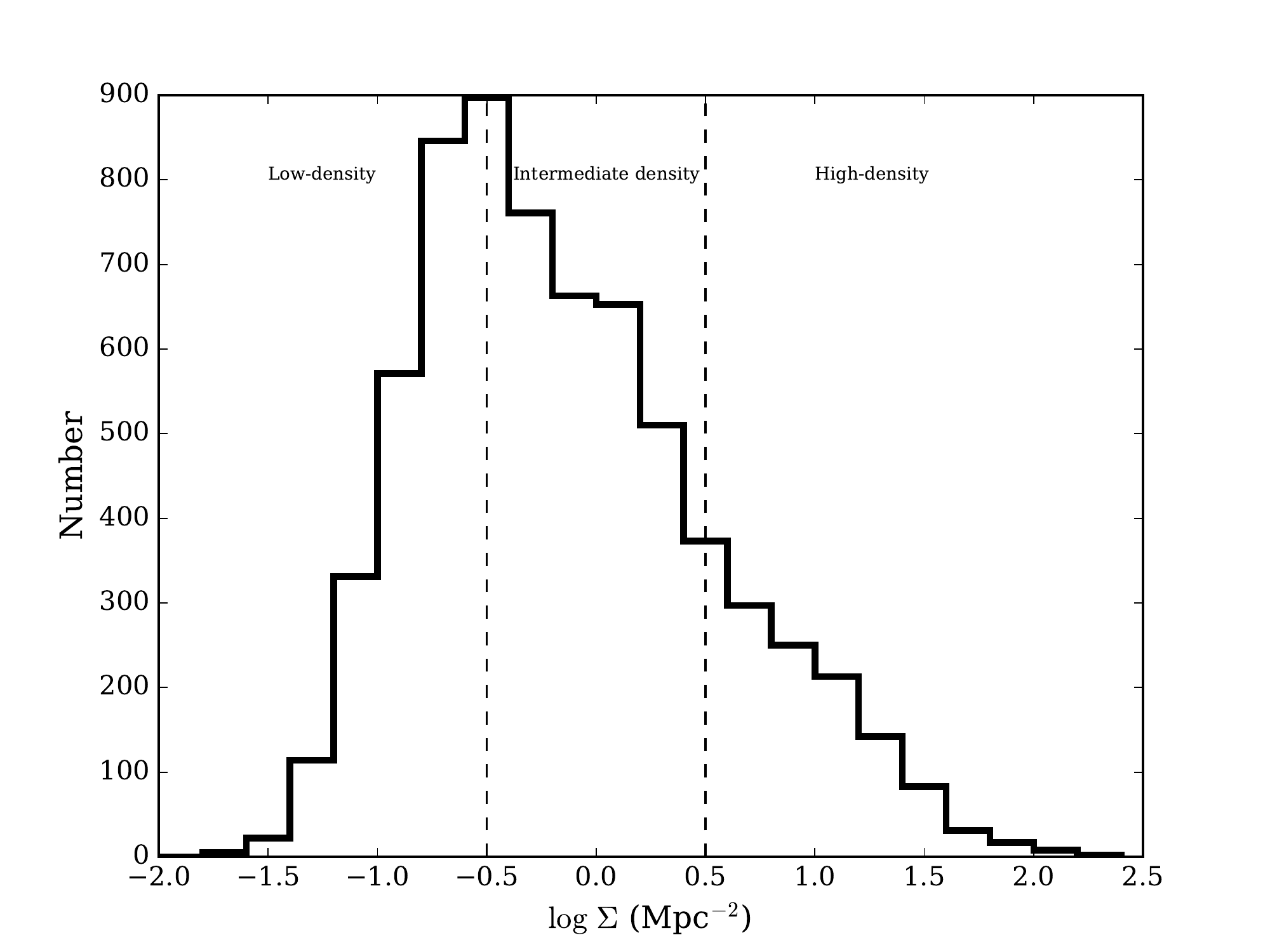}
\caption{Histogram of local environmental density ($\log \ \Sigma \ (\text{Mpc}^{-2})$) for our final sample of galaxies. The dashed lines separate the sample into low, intermediate, and high densities. } \label{density_hist}
\end{center}
\end{figure}

We use the local surface galaxy density from \citet{Baldry2006} as a measure of the environment. It is determined using, 
$\Sigma_\text{N} = \text{N}/\pi d_\text{N}^2$, where $d_\text{N}$ is the distance to the Nth nearest neighbour which are within the redshift range $\pm \Delta zc = 1000\ \text{km/s}$ for galaxies with spectroscopic redshifts or within the 95\% confidence limit for galaxies with photometric redshifts only. Following \citet{Baldry2006}, we use the best estimate, $\Sigma$, obtained by averaging the $\Sigma_N$ for the 4th and 5th nearest neighbour. Figure \ref{density_hist} shows the histogram of $\log \ \Sigma$ for our final sample of galaxies. Further we split our sample into low density ($\log \Sigma \ (\text{Mpc}^{-2}) < -0.5$),  intermediate density (-0.5 < $\log \Sigma \ (\text{Mpc}^{-2}) < 0.5$), and high densities ($\log \Sigma \ (\text{Mpc}^{-2}) > 0.5$).

We provide the MAGPHYS output along with information on morphology,
and environmental density from NA10, for our sample of galaxies in the
form of a catalogue. Table \ref{data table} shows a sub-sample of the
catalogue; the full table is available in the electronic version of
this paper.

\begin{table*} 
\caption{A part of the catalogue of the MAGPHYS output parameters and NA10 morphologies and  environmental density for each galaxy in our sample. The full catalogue is available in the electronic version of this paper.} \label{data table}
\begin{center}
\begin{tabular}{ccccccccc}
\hline \\
ObjID & R.A. & Decl. & z & $\log\ M_*$ & $\log \ sSFR$ & TT & $\log \ \Sigma$ & $\chi^2$ \\
& J2000 (deg) & J2000 (deg) & \ & (M$_{\odot}$) &  yr$^{-1}$& \  & (Mpc$^2$) &  \\   
\hline \\

J152757.03+032226.92 & 231.988 & 3.37414 & 0.085 & 11.71 & -12.72 & -5 & 1.199 & 3.078 \\
J152926.94+032851.96 & 232.362 & 3.4811 & 0.037 & 10.69 & -12.67 & -2 & 0.611 & 1.239 \\
J094953.04+001854.14 & 147.471 & 0.315039 & 0.064 & 11.2 & -10.62 & 4 & -0.507 & 2.568 \\
J014921.15+124254.20 & 27.3381 & 12.7151 & 0.034 & 10.45 & -10.32 & 5 & -0.75 & 0.847 \\
J083114.52+524225.06 & 127.811 & 52.707 & 0.064 & 11.03 & -10.57 & 2 & -0.699 & 0.554 \\
J085534.67+561207.52 & 133.894 & 56.2021 & 0.045 & 11.04 & -12.97 & -2 & -0.349 & 2.461 \\
J084714.08+012144.64 & 131.809 & 1.3624 & 0.04 & 10.9 & -12.17 & -5 & -0.625 & 0.818 \\
J120120.28+023125.91 & 180.335 & 2.52386 & 0.021 & 9.852 & -10.52 & 1 & -0.236 & 1.022 \\
J142720.13+025018.20 & 216.834 & 2.83839 & 0.027 & 10.54 & -10.27 & 5 & -0.13 & 0.232 \\
J092122.09+545154.28 & 140.342 & 54.8651 & 0.045 & 11.23 & -12.82 & -5 & 0.901 & 2.485 \\
\hline \\
\end{tabular}
\end{center}
\end{table*}

\section{Results and Discussion} \label{results}

Galaxy morphology, sSFR, and environmental density are correlated with
each other. It is possible that the correlation between any two of the
parameters is driven by the dependence on the third
parameter. Hence, in each of the following subsections we study the
correlation between any two of the parameters along with the
differential effect of the third parameter. We also compute the partial
correlation coefficient for each the three correlations while keeping
the third parameter as the control variable.

\subsection{Relation between morphology and specific star-formation rate}\label{morphology-sSFR}
\begin{figure*}
\centering
\includegraphics[scale=0.45]{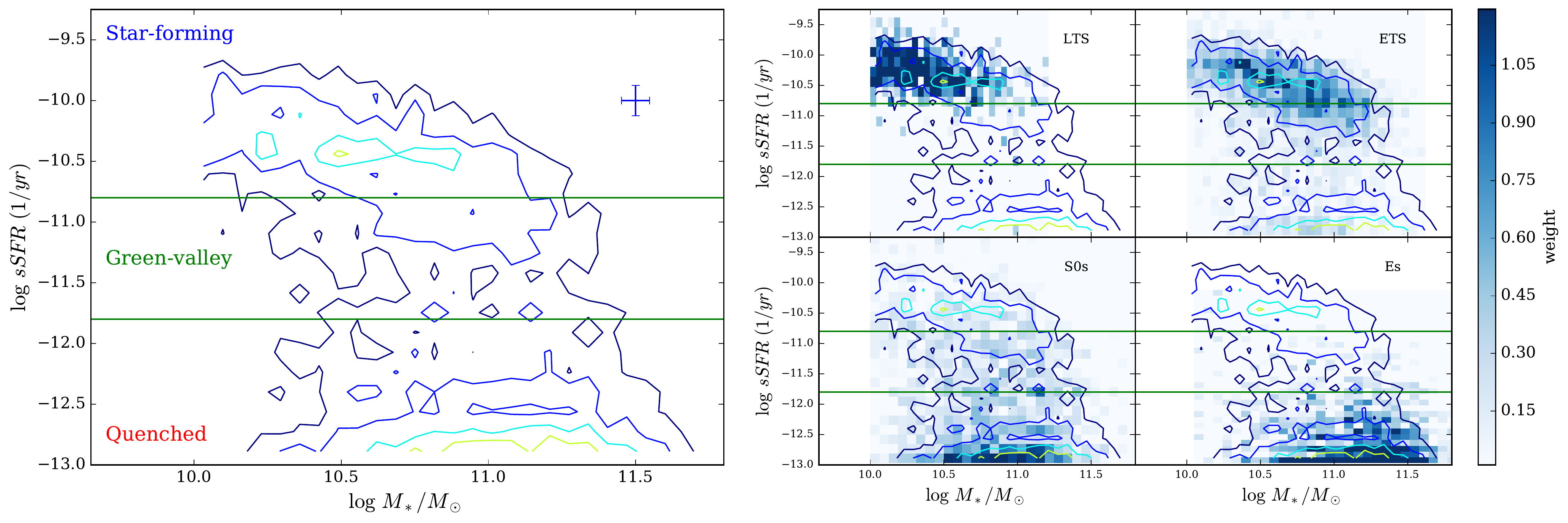}
\caption{Left Panel: Dependence of specific star formation rate on stellar mass for our entire sample. The contours shows the 1/V$_{max}$-corrected density of galaxies on the stellar mass and sSFR plane. The errorbars on the top-right corner shows the typical uncertanties in the estimating the stellar mass and sSFR from MAGPHYS. The region between the horizontal green lines defines the green valley from \citet{Salim2014}. Galaxies above the upper green line ($\log \text{sSFR} = -10.8$) are termed as star forming, and those lying under the lower green line ($\log \text{sSFR} = -11.8$) are termed as quenched galaxies. This shows the star-forming main sequence of galaxies, the quenched region at \stmg{10.5}, and a well seperated transition zone (green-valley) between the two populations. Right Panel: Same as the left-panel after splitting our whole sample, where top left: LTS, top right: ETS, bottom-left: S0s, bottom-right: Es. The shading in blue shows the 1/V$_{max}$-corrected density of galaxies in each morphological class. The contours are from the left panel for the full sample. We see a morphological dependance of the distribution of galaxies on sSFR-M$_*$ plane where the LTS are in the lower mass end (\stml{10.5}) star-forming sequence, ETS are in the high mass end of the star-forming sequence and near the green valley, S0s have a large population in the quenched region and a tail in the green valley, and the Es are mostly quenched.  }\label{green_valley}
\end{figure*}

\begin{table}
\begin{center}
\caption{Morphological fraction of galaxies which are in star-forming, green valley and quenched regions.}

\begin{tabular}{cccc}
\hline
 & Star-forming & Green-valley & Quenched \\
\hline
Total No. & 2184 & 1288 & 2789 \\
\hline
Es & 0.3\% & 8.3\% & 46.2\% \\
\hline
S0s & 11.8\% & 39.9\% & 42.1\% \\
\hline
Sa & 10.7\% & 18.4\% & 7.8\% \\
\hline
Sab & 11.0\% & 11.2\% & 1.3\% \\
\hline
Sb & 26.8\% & 15.1\% & 2.3\% \\
\hline
Sbc & 18.6\% & 4.4\% & 0.3\% \\
\hline
Sc-Sd & 20.8\% & 2.7\% & 0.0\% \\
\hline
\end{tabular} \label{distribution_of_sSFR}
\end{center}
\end{table}

\begin{figure*}
\includegraphics[scale=0.84]{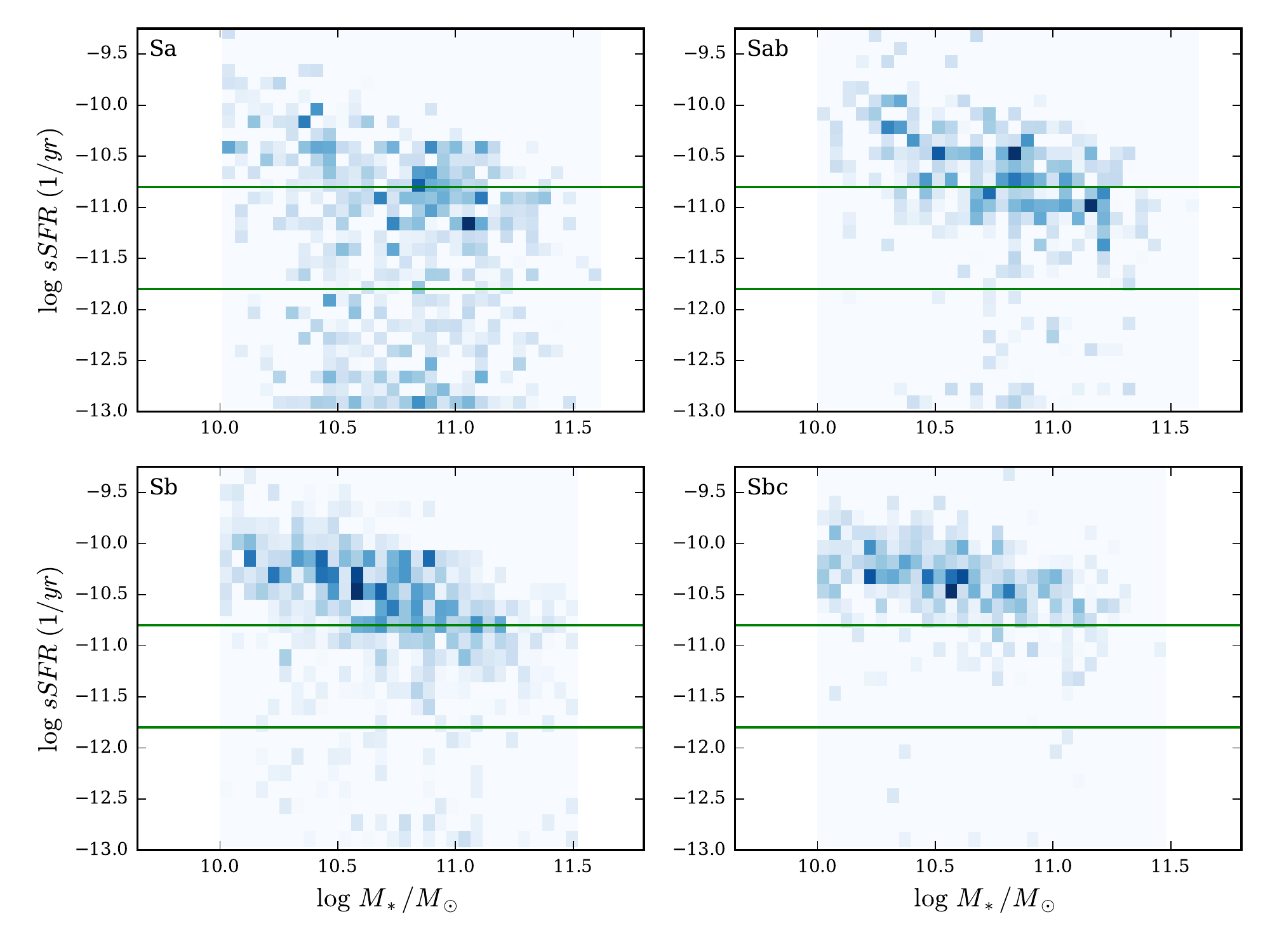}
\caption{Dependence of specific star formation rate on stellar mass for ETS further split into more detailed morphologies. The region between the horizontal green lines is the green valley. The density map shows the location of galaxies with different morphologies. Top left: Sa spirals, Top right: Sab spirals, Bottom-left: Sb, Bottom-right: Sbc. As we move along the Hubble sequence from Sa to Sbc, galaxies move from the quenched region through the green valley to the star-forming region.} \label{green_valley_esp}
\end{figure*}

In this section, we study the dependence of morphology on the
sSFR-M$_*$ plane. We then examine the fraction of galaxies of
different morphologies at fixed sSFR, and also the differential effect
of environment on it.

\subsubsection{Dependence of morphology on the sSFR-M$_*$ plane} \label{morphology sSFR Mstar plane}

We plot sSFR and stellar mass (M$_*$) for our sample in the left panel of Figure \ref{green_valley}. Most of the galaxies in our sample have a redshift below 0.07, and hence mass-incompleteness is not a major issue. We nevertheless correct for the mass-incompleteness in the sample using the 1/$V_{max}$ method \citep{Schmidt1968}. In this method we simply weight each galaxy by the maximum volume it can be detected referred by V$_{max}$. The weight for each galaxy is then 1/V$_{max}$. We use the 1/V$_{max}$ provided by the NA10 catalogue for every galaxy in our sample. In the left panel of Figure \ref{green_valley} the contours shows the 1/V$_{max}$-corrected density of galaxies on the stellar mass and sSFR plane. On the top-right corner the error bars shows the typical uncertainties in the estimation of stellar mass and sSFR. In order to estimate the typical uncertainity, we first find the median value of both the 16th and 84th percentile from the marginalized posterior probability distribution function for each parameter. And then take the mean of these two values which represent the error bar. The typical uncertainity in $\log \ \text{M}_*$ is about 0.048 dex, and in \sSFR\ is about 0.125 dex.

Following \citet{Salim2014}, galaxies above the upper green line ($\log \text{sSFR} = -10.8$) are termed as star forming galaxies (SFGs), and those lying under the lower green line ($\log \text{sSFR} = -11.8$) are termed as quenched galaxies (QGs). The region between the horizontal green lines in Figure \ref{green_valley} defines the green valley. We clearly find a bi-modality in the distribution of sSFR also previously noticed in the literature \citep{Brinchmann2004,Salim2007, Peng2010, Weinmann2006, Wetzel2012}.

We then separate our sample by morphology in to four groups: late-type
spirals (LTS), early-type spirals (ETS), S0s, and Ellipticals (Es). In
LTS, we group together Sc, Scd, and Sd morphologies, and in ETS we
group together Sa, Sab, Sb, and Sbc morphologies. We then plot each of
these groups separately on the sSFR-M$_*$ plot as shown in right panel
of Figure \ref{green_valley} in shades of blue. In this
  figure, the 1/V$_{max}$-corrected density map shows the position of
  a particular group on the sSFR-M$_*$ plane and the contours are
  repeated for the full sample from the left panel of Figure
  \ref{green_valley}. The region between the green lines is the green
valley. The star forming sequence is populated mostly by LTS at the
low mass end (\stml{10.5}) and by the ETS at the high mass end. ETS
with high mass are also present in the green valley and in the
quenched region.  Es and S0s populate the region defined for quenched
galaxies in the sSFR-M$_*$ plane with the S0s having a tail in the green
valley. Table \ref{distribution_of_sSFR} shows the percentage of
galaxies of a given morphology in the star-forming, green valley and
quenched regions. As expected, the star-forming region contains a very
small number of Es (0.3\%) and a small number of S0s (11.8\%). ETS,
comprising Sa, Sab, Sb, and Sbc morphologies, constitute 67.1\% and
LTS constitute 20.1\% of the galaxies in the star-forming
region. Thus, in the local Universe majority of the recent star formation in
massive galaxies happens in ETS and LTS. On the other hand, the
quenched region predominantly contains Es (46.2\%) and S0s (42.1\%)
with a small contribution from ETS (11.7\%) and almost no contribution from LTS. In the green valley, majority is populated by ETS
(49.1\%) and S0s (39.9\%) and a small percent are Es (8.3\%) and LTS
(2.7\%). \citet{Bouquin2015} also found that in the green valley,
defined using GALEX data, S0-Sa galaxies are more common. We find
that, in the local Universe, massive galaxies in the green valley are
predominantly disk galaxies with a prominent bulge. Further, similar
results were also seen from the Galaxy Zoo (GZ) project by
\citet{Schawinski2014} which identified galaxies as only early-types
and late-types. Since we have detailed morphological classification, we can have a closer look at the morphology of galaxies undergoing quenching. In particular, a comparison of Fig. \ref{green_valley} with \citep[][ see Fig. 3]{Schawinski2014} shows that for massive galaxies, most of the GZ late-type galaxies in the green valley and quenched region are likely to have Sa-Sb morphologies. The small number of GZ early-type galaxies in the green valley are mostly S0 galaxies.

In Figure \ref{green_valley_esp}, we split the ETS into their morphological subtypes to explore their distribution on the sSFR-M$_*$ plane. We can see that as the bulge becomes more prominent when we move along Sbc to Sa morphology, galaxies gradually make a transition to the quenched region through the green valley. Table \ref{distribution_of_sSFR} also shows that the star-forming region comprises higher percentage of Sbc and Sb galaxies than Sab and Sa. Whereas in the green valley there is a higher percentage of Sa, Sab, and Sb galaxies relative to Sbc galaxies. Also, in the quenched region Sas galaxies are more numerous than Sab, Sb, and Sbcs. The main systemic difference between these different subtypes of ETS is in the size and luminosity of the bulge component. The variation we see points towards the importance of the bulge component in the quenching of galaxies. A similar result was also found by \citet{Bluck2014}, wherein they found that the bulge mass correlates most with the passive state of a galaxy. \citet{Mendez2011} studied the morphologies of green valley galaxies in the redshift range $0.4 < z < 1.2$ and found that 14\% are in merger, 51\% are late-type galaxies, and 35\% are early-type galaxies. Under the simplistic assumption that all of the merging galaxies in that redshift range will eventually transform into early-type galaxies, the fraction of green valley early-type galaxies will increase to 49\%, which is very close to the fraction of early-type galaxies (Es and S0s) in the green valley that we find, which is  48.1\%. In reality, the situation may be much more complicated wherein the environment  also plays a role in transforming late-type galaxies into early-type galaxies.

\subsubsection{Fraction of galaxies of different morphologies at fixed sSFR, and the differential effect of environment} \label{fraction morph fixed sSFR}

\begin{figure*}
\centering
\includegraphics[scale=0.9]{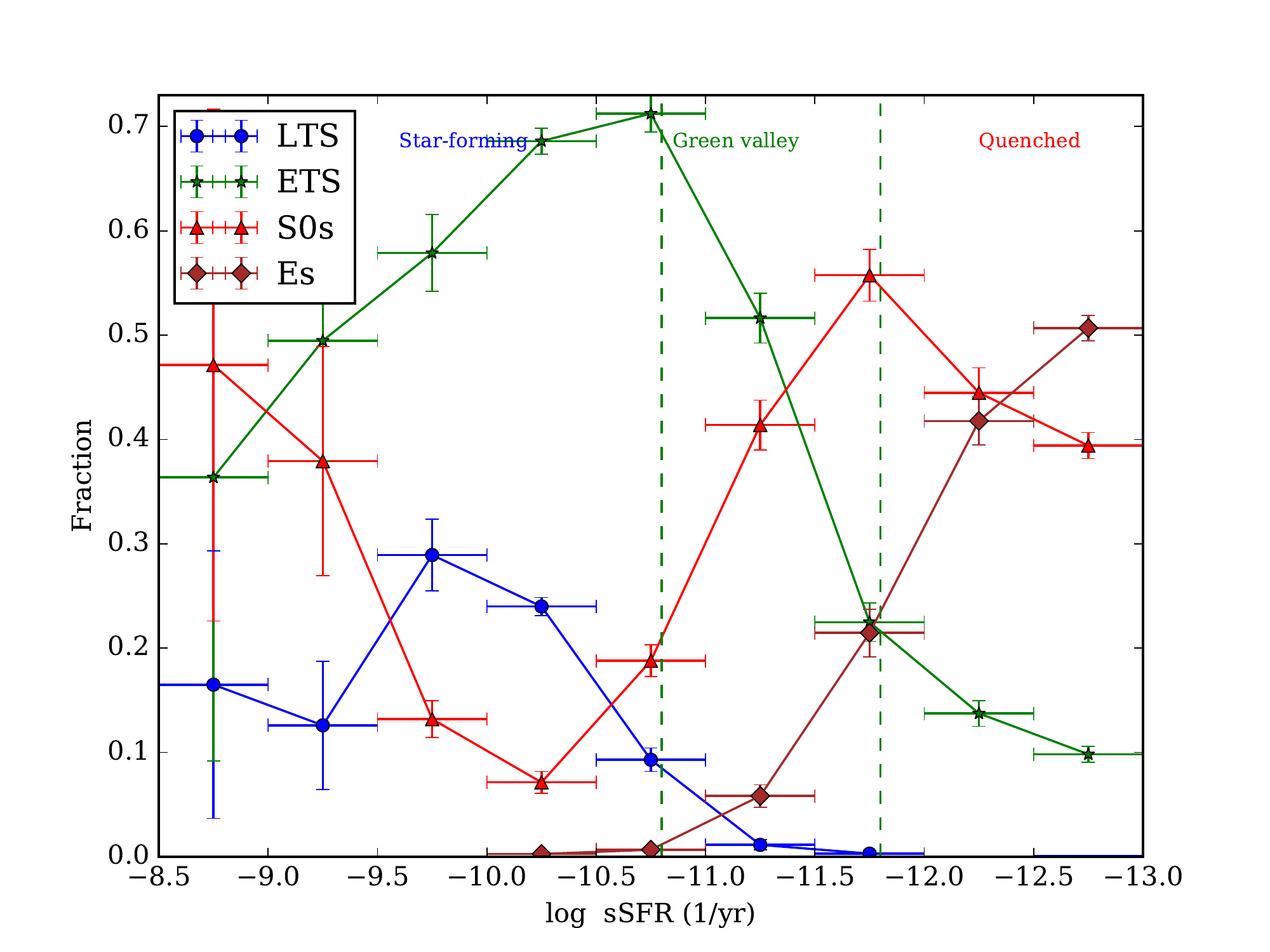}
\caption{Fraction of galaxies of different morphological type (LTS in blue, ETS in green, S0s in red, and Es in brown) as a function of \sSFR. The errorbars are jacknife errors. Fraction of LTS and ETS decline with decreasing \sSFR, which coincides with an increase in the fraction of S0s and Es. S0s on the other hand also show a second population, which is actively forming stars. See text for a detailed discussion.} \label{morph-sSFR-frac}
\end{figure*}

\begin{figure*}
\centering
\includegraphics[scale=0.5]{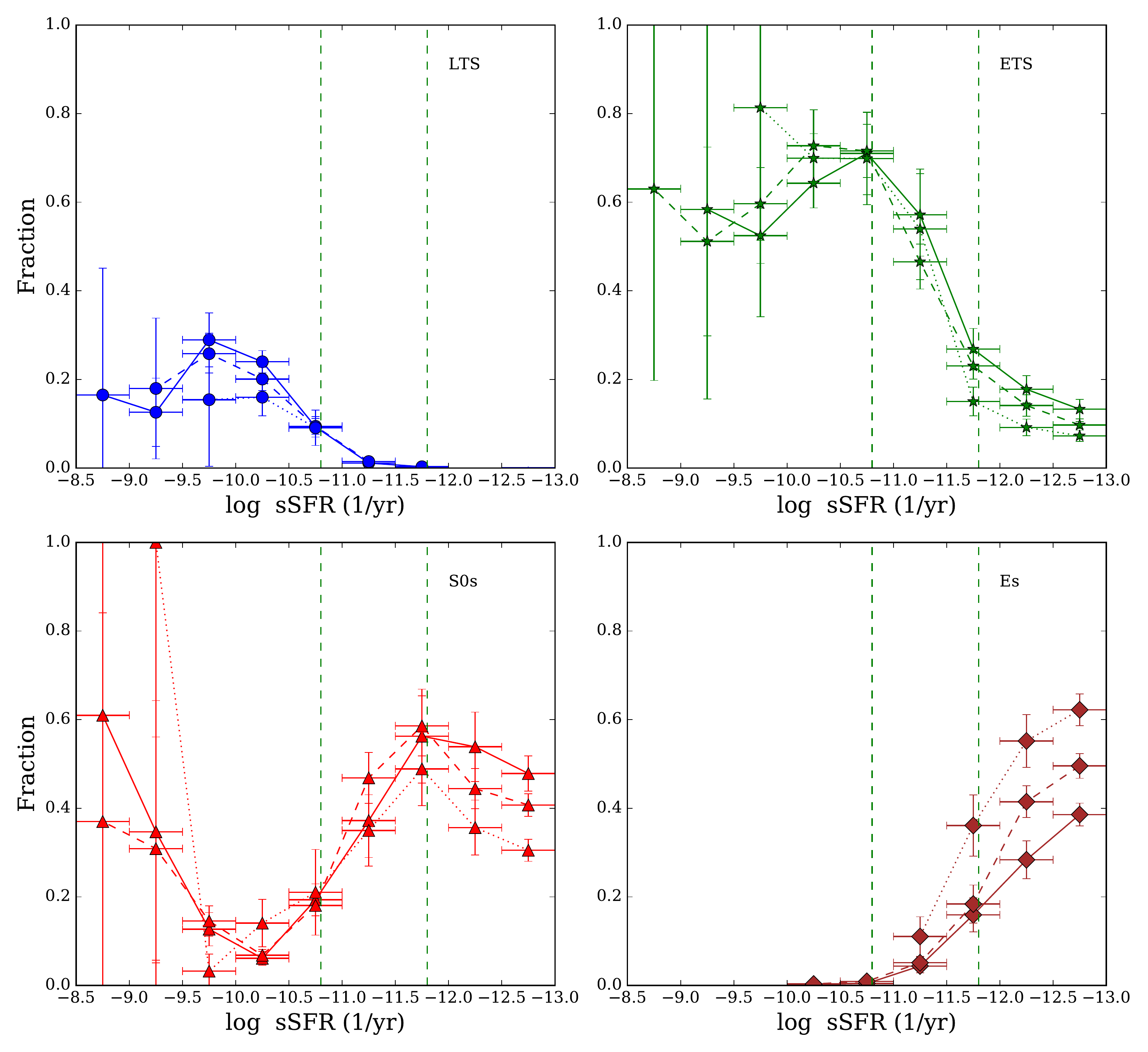}
\caption{Fraction of galaxies of different morphological type (the colour code is same as Fig. \ref{morph-sSFR-frac}) in each panel. And in each panel we further split with environmental density in terms of low density ($\log\Sigma \leq -0.5$)-filled line, intermediate ($-0.5 <\log\Sigma < 0.5$) density-dashed line, and high density ($\log\Sigma \geq 0.5$)-dotted line. The errorbars are jacknife errors. Notice that fraction of ETS at fixed \sSFR\ does not change significantly with environment. Es are more abundant in intermediate and high densities. LTS are more abundant in low and intermediate densities. There continue to be two populations of S0s (green valley/quenched S0s and star forming S0s) in all environements.}\label{morph-sSFR-fraction-diff1}
\end{figure*}

We study the fraction of LTS, ETS, S0s, and Es in bins of $\log$ sSFR
(Fig \ref{morph-sSFR-frac}). Here again we correct for the Malmquist bias, using the 1/V$_{max}$ method, while calculating the fractions. In each of the \sSFR\ bin, we find the fraction of each morphological type as follows \citep[cf. ][eq. 2]{vandenBosch2008}:
\begin{equation}
f_{\text{morph}| sSFR} = {\sum\limits_{i=1}^{i=N_a}w_i}\bigg /{\sum\limits_{j=1}^{j=N_b}w_j},
\end{equation} 
where $w_i$ is the weight for each galaxy calculated using 1/V$_{max}$. And $N_a$ is the total number of galaxies of a particular morphlogical type for a given \sSFR\ bin. $N_b$ is the total number of galaxies in that \sSFR\ bin. In all the sections that follow, all the fractions are 1/V$_{max}$-corrected using this method. The errors on these fractions are calculated using the jacknife technique following the approach described in \citet{vandenBosch2008}. 

It is interesting to see in Fig. \ref{morph-sSFR-frac} that
  the trends for different morphological types are quite different. In
  accord with earlier results, LTS have a higher fraction ($\sim 0.3$)
  for \sSFR\  $\sim -10$ and there is a sharp decline in the fraction of
  LTS with decreasing \sSFR. Meanwhile, the fraction of ETS shows a
steep increase as \sSFR\ decreases, with a peak at the boundary of the
star-forming and green valley regions. This is followed by a sharp
decline in the green valley and then settling at about 10\% in the
quenched region. Such a decline in the ETS and LTS coincides with an
sharp increase in the fraction of S0s and Es with decreasing \sSFR. A
simple interpretation is a morphological transformation of LTS and ETS
to S0s and Es, which is accompanied by a decline in the sSFR.
Our results are in agreement with the GZ based study of galaxy
  morphology and color \citep[][see Fig. 5]{Skibba2009} where they see
  a correlation between morphology and color such that the likelihood
  of early-type galaxy increases as the galaxy color becomes
  redder. We can put a stronger constraint on the morphologies by
  suggesting that they are ETS transforming in S0 galaxies in the
  green valley. Several physical processes can lead to such
  morphological transformations from late-type to early-type galaxies
  which also quench the star formation in these galaxies
  \citep[see][for a review and the references
    therein]{Boselli2006}. This has also been observed in high
  redshift, upto $\sim 1$ \citep{Kovac2010}. Interestingly, there are
also two distinct class of S0s galaxies: one which is in the green
valley and quenched region and thus hosting an older stellar
population, and the other which is actively forming stars (with \sSFR\
in the range [-9, -8]) hosting a younger stellar population. The
latter class of S0s are almost as abundant as the LTS in that \sSFR\
bin. \citet{Barway2013} have found similar populations of
  young and old S0s using UV-near IR colours of
  galaxies. \citet{Johnston2014} have found S0s in clusters with a
  younger bulge thus proposing that the final episode of star
  formation might be happening in the central regions of cluster S0s.

Some of the physical processes which lead to changes in the star
formation and morphologies of galaxies are thought to depend on
environment which hosts these galaxies, e.g. ram-pressure stripping \citep{Gunn1972}, galaxy harassment \citep{Farouki1981, Moore1996}, and strangulation \citep{Larson1980}. Hence, we need to we study the
impact of the environment on the relation between morphology and sSFR.

To do this, we split our sample into low density ($\log\Sigma \leq -0.5$), intermediate ($-0.5 <\log\Sigma < 0.5$) density, and high density ($\log\Sigma \geq 0.5$) environments and study the fraction of different morphologies as a function of  \sSFR\ (Fig. \ref{morph-sSFR-fraction-diff1}). We notice that environmental density does not significantly alter the fraction of individual morphologies as a function of \sSFR, except for Es, which are more abundant in intermediate and high density environments. This is a manifestation of the morphology-density relation \citep{Dressler1980}. The two populations of S0s seen in Figure \ref{morph-sSFR-frac} continue to be present in all the three environments. S0s in green valley and the quenched region show similar trends in all density environments, albeit with large uncertainities.  

This result, along with our analysis presented in Section \ref{partial corr coeff}, indicates that morphology is strongly correlated with sSFR, independent of the environment.

\subsubsection{Fraction of star-forming, green valley, and quenched galaxies at fixed morphological T-type, and the differential effect of environment} \label{fixed-morph}

\begin{table*}
\begin{center}
\caption{Fraction of star-forming, green valley, and quenched galaxies each morphological type.}

\begin{tabular}{ccccccccc}
\hline
 & E & S0 & Sa & Sab & Sb & Sbc & Sc-Sd \\ 
\hline
Total No. of galaxies & 1478 & 1787 & 671 & 429 & 846 & 502 & 548 \\
\hline
Star forming galaxies & 0.4\% & 13.5\% & 34.1\% & 56.9\% & 69.6\% & 86.1\% & 92.5\% \\ 

Green valley galaxies & 7.9\% & 26.9\% & 34.8\% & 34.4\% & 23.2\% & 12.1\% & 7.2\%\\ 

Quiescent galaxies  & 91.7 \% & 59.6\% & 31.1\% & 8.7\% & 7.2\% & 1.8\% & 0.3\% \\ 
\hline
\end{tabular} \label{sSFR-morph-table}
\end{center}
\end{table*}

\begin{figure}
\centering
\includegraphics[scale=0.45]{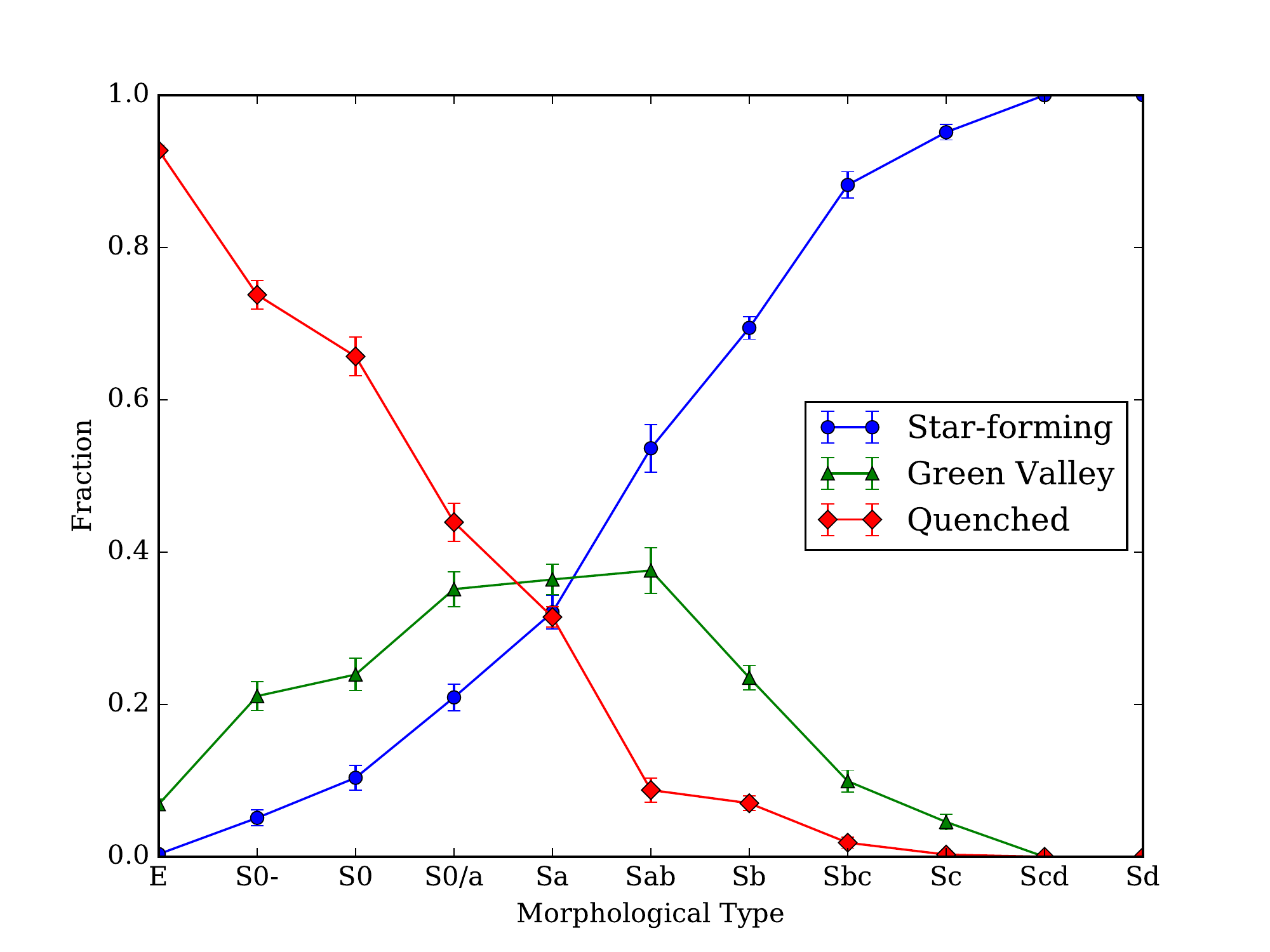}
\caption{Fraction of star-forming (blue), green valley (green) and quenched (red) galaxies for a fixed morphological T-type going to Es to Im spirals. The errorbars are jacknife errors. Notice that the quenched fraction decreases as we move from early-type galaxies to late-type. The fraction of green valley galaxies increases from Es to S0s and upto Sab spirals beyond which there is a steep decline. Interestingly Sa spirals shows equal fractions of all three stages of star-formation.} \label{sSFR-morph}
\end{figure}

\begin{figure*}
\centering
\includegraphics[scale=0.4]{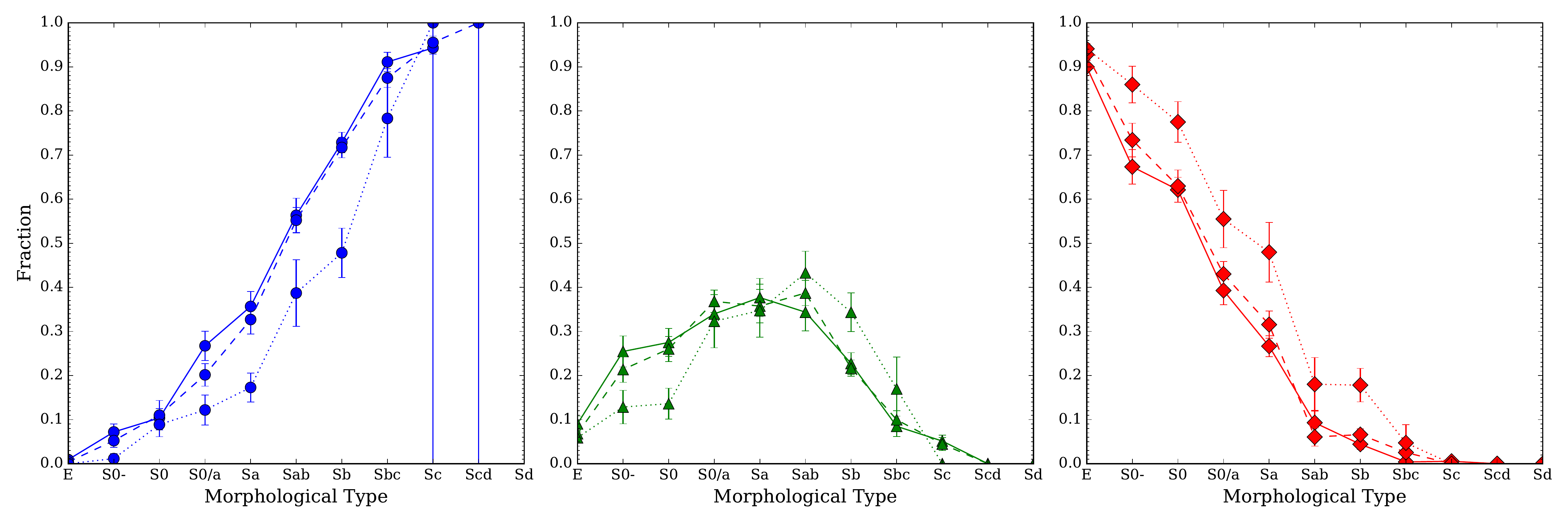}
\caption{Fraction of star-forming (blue), green valley (green) and quenched (red) galaxies for a fixed morphological T-type shown in the left, middle, and right panel respectively. In each panel we further split in three environmental bins: low density ($\log\Sigma \leq -0.5$)-filled line, intermediate ($-0.5 <\log\Sigma < 0.5$) density-dashed line, and high density ($\log\Sigma \geq 0.5$)-dotted line. The errorbars are jacknife errors. Notice that in all environments as we go from early-type galaxies to late-type the quenched fraction decreases. However, the effect of environment on quenching is most apparent on Sa and S0s, wherein their quenched fraction increases susbtantially in high density environments. Interestingly, in low density environments the fraction of star-forming S0/a galaxies is high.}\label{sSFR-morph-fraction-diff2}
\end{figure*}

We divide our whole sample of galaxies into three parts as star-forming (\sSFR\ $\geq -10.8$), green valley ($-10.8 > $ \sSFR\ $> -11.8$) and quenched (\sSFR\ $\leq -11.8$). We then plot the 1/V$_{max}$-corrected fraction of galaxies in each of these types for a fixed morphological T-type (Fig. \ref{sSFR-morph}) using the same method illustrated in the earlier sub-section. As we use all the morphological subtypes in this section, we have S0s in three stages: \som (S0s with a smooth light profile), S0 (S0s with some structure in their light profile), and S0/a (S0s in a transition from Sa spirals). Spiral galaxies range from Sa to Sd.

The overall trend seen in Figure \ref{sSFR-morph}, that as we go from
early-type to late-type galaxies the quenched fraction decreases, is
well expected. However, it is interesting to see that some of the Es
are in the green valley (7.9\%), and a negligible fraction of them are
star-forming (0.4\%). These Es may have recently undergone gas-rich
minor/major merger which leads to trigerring of star-formation in these
galaxies. About 70\% of the \som and 65\% of the S0 galaxies are
quenched. Thereafter, there is a rapid fall in the fraction of
quenched S0/a galaxies (and hence a corresponding increase in the
star-forming and green valley S0/a). Further, the quenched fraction of
Sa spirals only slightly decreases, and interestingly it is almost
equal to both the fraction of green valley and star-forming Sa
spirals. Spirals from Sab to Sd show a rapid decline in the quenched
fraction and green valley. These fractions in terms of percentages are
summarized in Table \ref{sSFR-morph-table}.

We also examine the role of environment in the quenched fraction of a
fixed morphological type. We plot the fraction of star-forming, green valley, and quenched galaxies
for a fixed morphological T-type in three different panels. In each panel, 
we further split in terms of three
environmental bins: low, intermediate, and high density (Fig \ref{sSFR-morph-fraction-diff2}). The quenched fraction of Es in all the three
environments remains high and almost the same with a small fraction of
them in the green valley. LTS (Sc-Sd) remain mostly star forming
in all environments.

Interestingly, quenched ETS are present in all environments and their
quenched fraction increases with increasing environment density. The
strongest effect of environment is seen on the Sa spirals and S0/a
galaxies, wherein their quenched fraction increases with increasing
environmental density. Note that the fraction of quenched Sa spirals which is around 30\% in low and intermediate density environments, suddenly
rises to about 50\% in the high density environment. This shows that
environment does play a role in quenching of Sa spirals, however the
physical process of quenching is likely to be gentle since their
spiral arms are intact. This rules out ram-pressure stripping for the
quenching of these galaxies, and makes a gentler process like
strangulation more likely. Our results are in agreement with those of \cite{vandenBosch2008} and \cite{Peng2015}, where the authors argue for strangulation as the dominant quenching mechanism in local galaxies.

The quenched fraction of S0/a galaxies, which are in transition from Sa
to S0, also increases with increasing environmental density.  \som and
S0 are mostly quenched in all environments with their fraction
increasing by about 10\% in high density environment. The abundance of
quenched \som and S0 in intermediate and particularly in low density environments is intriguing. According to \cite{Baldry2006} (see their Fig. 9 caption), $\log \Sigma > 0.8 \ \text{Mpc}^{-2}$ corresponds to
cluster like environments. Thus, our definition of low and
intermediate densities, $\log \Sigma < 0.5\  \text{Mpc}^{-2}$,
will correspond to field and group like environment. Thus, at low densities, ram-pressure stripping
may not be a dominant mechanism in the formation of S0 galaxies. Interestingly, in low density environments
about 30\% of the S0/a galaxies are star-forming. Even though this
fraction decreases with increasing density, in intermediate and high
densities about 20\% and 10\% of the S0/a galaxies are
star-forming, respectively. About 10\% of the S0 and \som are star-forming in
the low density environment, and their fraction decreases
when we go to the high density environment. The presence of such
star-forming S0s in all environments gives rise to the second
population of star-forming S0s previously seen in Figure
\ref{morph-sSFR-fraction-diff1}.

Thus, we conclude that for massive galaxies in the local Universe,
only the quenched fraction of S0/a and S0 galaxies show a significant
dependence on environmental density. Further, the fact that the
quenched fraction of S0/a galaxies is more close to that of Sa spirals and
much lower than that of S0 and \som, shows that S0/a galaxies likely undergo
a different quenching process than that of S0 and \som galaxies.

\subsection{Relation between morphology and local environment} \label{morphology environment}
\begin{figure}
\centering
\includegraphics[scale=0.45]{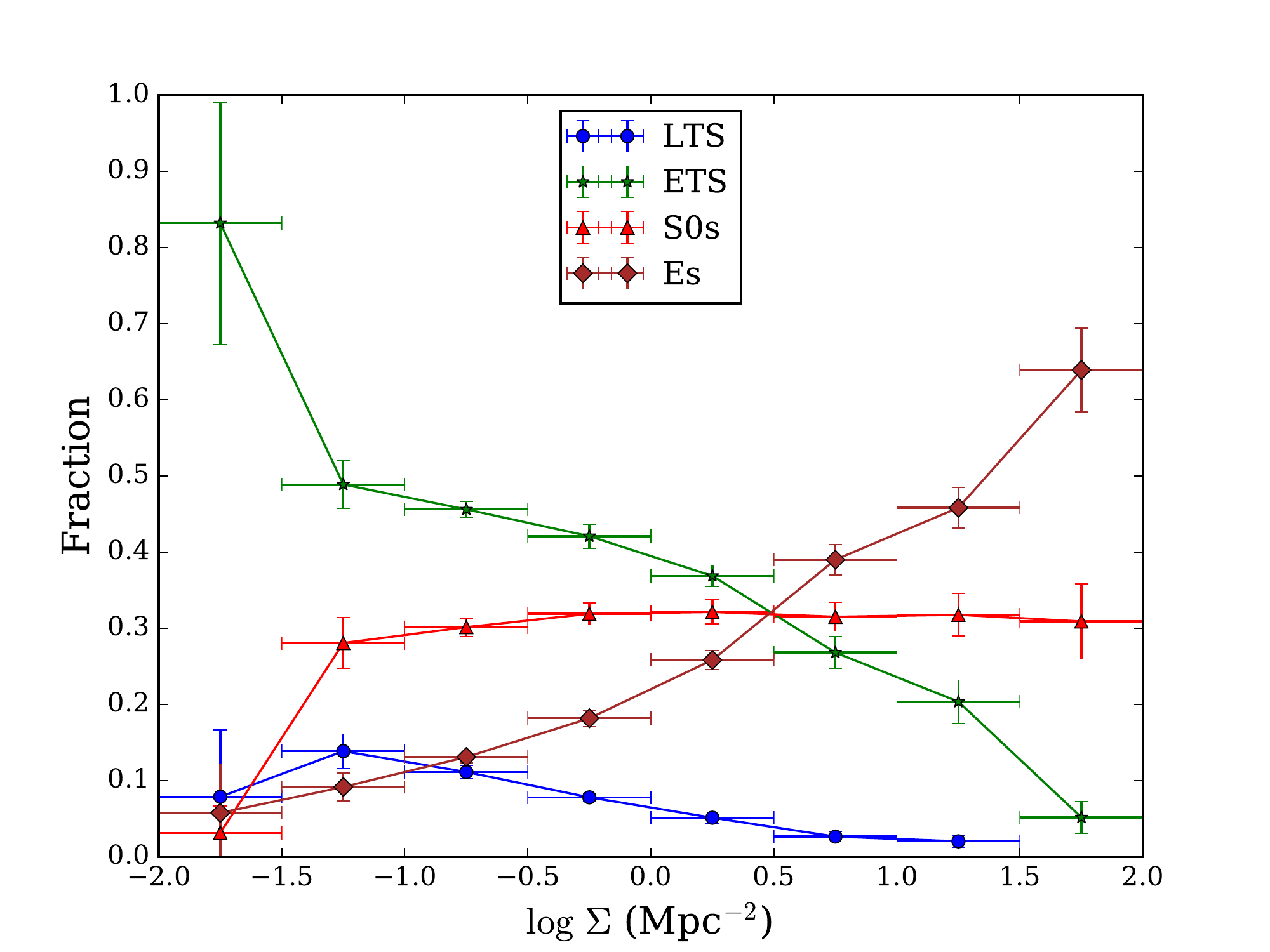}
\caption{Fraction of LTS (blue), ETS (green), S0s (red), and Es (brown) in fixed environmental bins. The errors shown are jacknife errors. Notice the fraction of ETS decreases, and the fraction of Es increases in the high density environments.} \label{morph-density}
\end{figure}

\begin{figure*}
\centering
\includegraphics[scale=0.4]{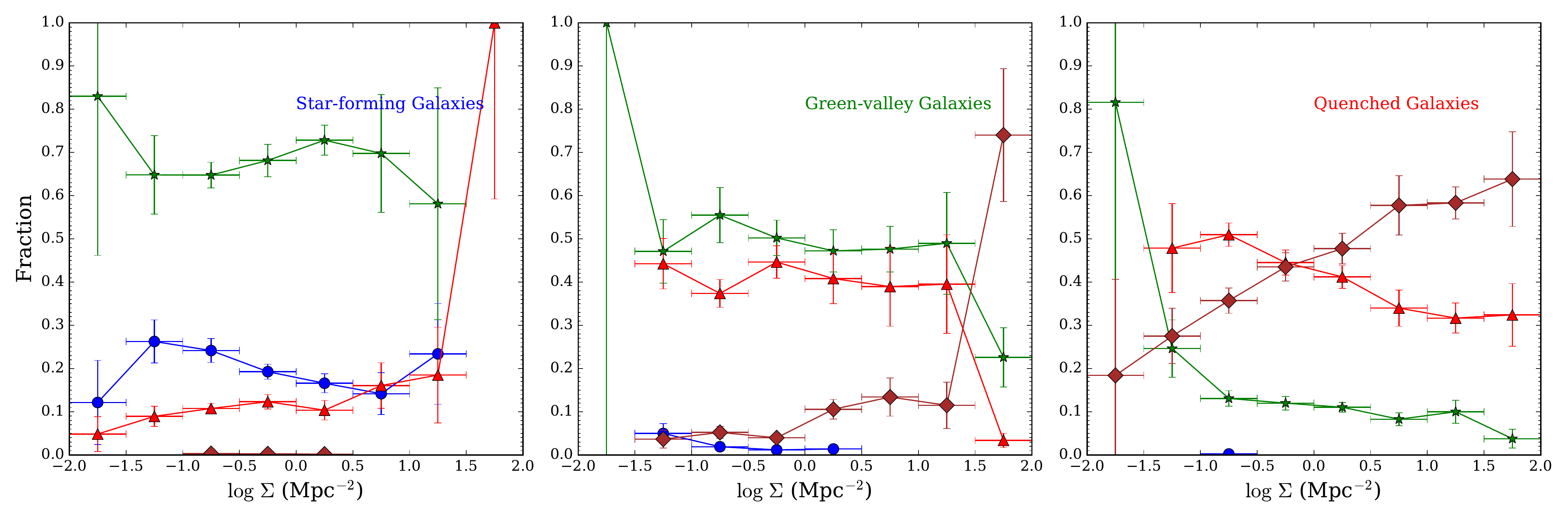}
\caption{Fraction of LTS (blue), ETS (green), S0s (red), and Es (brown) in fixed environmental bins split in three parts: star-forming galaxies in the left panel, green valley galaxies in the middle panel, and quenched galaxies in the right panel. The errors shown are jacknife errors. Notice the presence of star forming S0s at all environmental densities. ETS and S0s populate the green valley galaxies in all environments, except in the highest environmental density bin. Further, the fraction of green Es increase with increasing environmental density. See the text for a detailed discussion. }\label{morph-density-diff1}
\end{figure*}

\begin{table}
\begin{center}
\caption{Fraction of galaxies of different morphological types in low, intermediate, and high density environments.}

\begin{tabular}{cccc}
\hline
 & Low density & Intermediate density & High density \\
\hline
Total No. & 2158 & 2968 & 1135 \\
\hline
Es & 12.0\% & 21.4\% & 42.7\% \\
\hline
S0s & 29.2\% & 31.9\% & 31.4\% \\
\hline
Sa & 13.4\% & 10.9\% & 7.1\% \\
\hline
Sab & 7.9\% & 6.9\% & 4.5\% \\
\hline
Sb & 15.7\% & 14.1\% & 8.4\% \\
\hline
Sbc & 9.4\% & 7.9\% & 3.5\% \\
\hline
Sc-Sd & 12.4\% & 6.9\% & 2.4\% \\
\hline
\end{tabular} \label{distribution_of_sigma}
\end{center}
\end{table}

\begin{table*}
\begin{center}
\caption{Fraction of galaxies in low, intermediate, and high densities for different morphological types.}

\begin{tabular}{ccccccccc}
\hline
 & E & S0 & Sa & Sab & Sb & Sbc & Sc-Sd \\ 
\hline
Total No. of galaxies & 1478 & 1787 & 671 & 429 & 846 & 502 & 548 \\
\hline
Low density & 18.2\% & 31.8\% & 40.8\% & 38.9\% & 38.8\% & 41.3\% & 52.7\% \\ 

Intermediate density & 46.8\% & 48.3\% & 47.6\% & 49.0\% & 50.0\% & 50.3\% & 41.8\%\\ 

High density  & 35.0\% & 18.4\% & 11.6\% & 12.1\% & 11.2\% & 8.4\% & 5.5\% \\ 
\hline
\end{tabular} \label{density-morph-table}
\end{center}
\end{table*}

Next we study the dependence of morphology on the local environmental density. We follow a similar procedure as in Section \ref{morphology-sSFR}. We first examine the fraction of galaxies with different morphologies in fixed environmental bins.  We then study the differential effect of sSFR on this relation. 

In Figure \ref{morph-density}, we plot the the 1/V$_{max}$-corrected  fraction of Es, S0s, ETS, and LTS galaxies in fixed environmental bins. In high density
environments ($\log \Sigma > 0.5$) the fraction of ETS decreases and
the fraction of Es increases. This is the well known
morphology-density relation \citep{Dressler1980}. Interestingly, the
fraction of S0s does not increase as we go to high densities. One
reason might be that we only have massive galaxies in our sample, and
the effects of environment (e.g. ram-pressure stripping), which can
transform a spiral into an S0, is stronger for low-mass galaxies \citep{Mori2000, Marcolini2003, Bekki2009, Fillingham2016, Emerick2016}. LTS are most abundant in very low densities ($\log \Sigma \sim -1.5$) and
gradually decline with increasing environmental density.  The
morphological fraction of galaxies split in three environmental bins:
low-density, intermediate density, and high density is summarised in
Table \ref{distribution_of_sigma}.

We split our sample into  star-forming, green valley, and quenched galaxies and plot the fraction of galaxies with different morphologies in fixed environmental bins (Fig. \ref{morph-density-diff1}). Notice that the morphology-density relation does change, this is partly because of the paucity of star-forming Es and quenched/green valley LTS. In the left panel of Fig. \ref{morph-density-diff1}, fraction of star-forming LTS gradually declines with increasing environment and the fraction of star-forming ETS increases, although within errorbars it is not very significant. Except that in the highest density bin ($1.5 < \log \Sigma \leq 2.0$) the trend is reversed; however this feature is not very significant due to the large errorbars. Interestingly, in all environments, except in the lowest density bin,  about 10\% of the star-forming galaxies are S0s. These are the second population of actively star-forming S0s observed in Figure \ref{morph-sSFR-fraction-diff1}. The ubiquity of these S0s in all environments, from low density to high density requires further study. In all environments, the green valley galaxies (middle panel of Fig. \ref{morph-density-diff1}) are mostly ETS (around 50\%) and S0s (around 40\%), except in the highest density bins which show a decline. The fraction of green valley Es increases with increasing density with a steep rise in the highest density bin, wherein around 75\% of the green valley galaxies are Es. These might be Es which have undergone recent gas rich minor/major merger(s). Since the rate of merger increases with increasing environmental density we find more Es in green valley in such environments. For quenched galaxies (right most panel of Fig. \ref{morph-density-diff1}), the fraction of Es increase with increasing environmental density which most likely arises from the morphology-density relation. Except in the lowest density bin, the fraction of quenched S0s show a weak decline with increasing density, which is possibly because of the increase in the fraction of Es. The fraction of ETS declines very rapidly from the lowest density bin towards intermediate densities and then remains constant at around 10\% with a small decline in the highest density bin. Interestingly, in the lowest density bin ($\log \Sigma < -1.5$) there are no quenched S0s, and about 80\% of the galaxies are ETS and about 20\% are Es. 

We examine the percentage of Es, S0s, ETS, and LTS in low, intermediate, and high densities as shown in Table \ref{density-morph-table}. Surprisingly, most of the Es, S0s, and ETS are present in the intermediate density environment.

\subsection{Relation between specific star formation rate and local environment} \label{sSFR environment}
\begin{figure}
\centering
\includegraphics[scale=0.45]{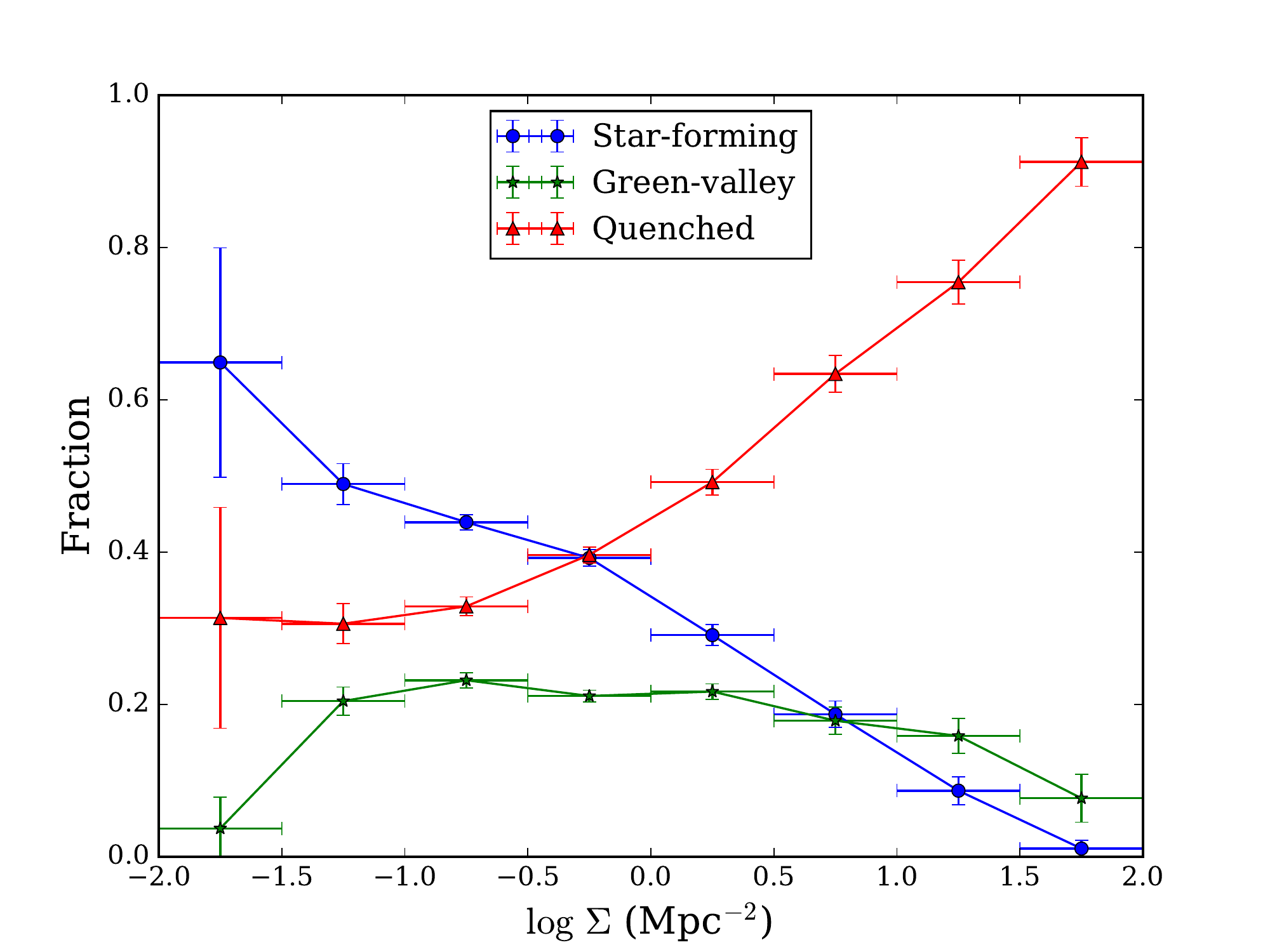}
\caption{Fraction of star-forming (blue), green valley (green), and quenched (red) galaxies for fixed environmental bins. The errorbars are jacknife errors. Notice that the fraction of star-forming galaxies decreases with increasing environmetal density. The fraction of green valley galaxies does not show a significant change with environmental density. } \label{density-sSFR}
\end{figure}

\begin{figure*}
\centering
\includegraphics[scale=0.55]{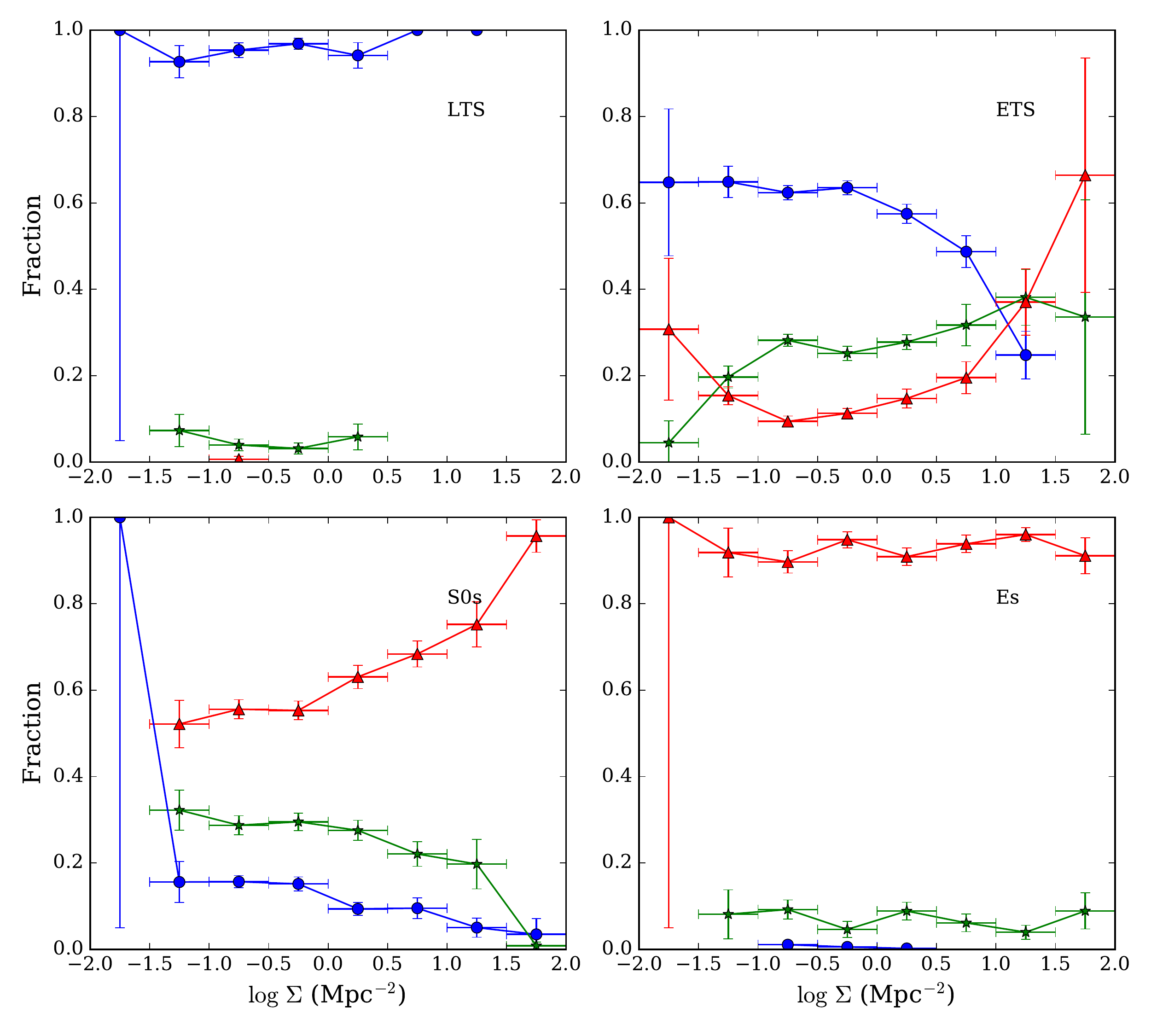}
\caption{Fraction of star-forming (blue), green valley (green), and quenched (red) galaxies for LTS (top left), ETS (top right), S0s (bottom left), and Es (bottom right). LTS(Es) are star-forming(quenched) as a function of environment. The errorbars are jacknife errors. Star-forming ETS fraction decreases and quenched S0 fraction increases with increasing environmental density.  }\label{density-sSFR-fraction-diff1}
\end{figure*}

\begin{table}
\begin{center}
\caption{Fraction of star-forming, green valley, and quenched galaxies in low, intermediate, and high densities.}

\begin{tabular}{cccc}
\hline
    & Low density & Intermediate density & High density\\ 
\hline
Total No. & 2158 & 2968 & 1135 \\
\hline
Star forming & 46.1\% & 35.1\% & 15.6\% \\ 

Green valley & 21.9\% & 21.6\% & 17.3\% \\ 

Quenched  & 32.0\% & 43.3\% & 67.1\% \\ 
\hline
\end{tabular} \label{density-sSFR-table}
\end{center}
\end{table}

\begin{table}
\begin{center}
\caption{Fraction of galaxies in low, intermediate, and high densities which are in star-forming, in green valley, and in the quenched regions.}

\begin{tabular}{cccc}
\hline
    & Star forming  & Green valley  & Quenched \\ 
\hline
Total No. & 2184 & 1288 & 2789 \\
\hline
Low density & 44.0\% & 35.1\% & 24.6\% \\ 

Intermediate density & 48.0\% & 49.9\% & 47.7\% \\ 

High density & 8.0\% & 15.0\% & 27.7\% \\ 
\hline
\end{tabular} \label{sSFR-density-table}
\end{center}
\end{table}

We now study the relation between the two remaining
parameters, sSFR and environment. We follow a similar procedure
as in the earlier two subsections. We first examine the fraction
of star-forming, green valley and quenched galaxies for fixed
environmental bins. Then we use the morphological classification
in terms of LTS, ETS, S0s and Es to study the differential effect of
morphology on the relation between sSFR and environment.

The 1/V$_{max}$-corrected  fraction of star-forming,
green valley, and quenched galaxies in fixed environmental
bins is shown in Figure \ref{density-sSFR}. The quenched fraction increases with increasing
environmental density (and hence the star-forming fraction
decreases). Interestingly, the fraction of green valley galaxies
remains almost constant at about 20\% in all environments, except in the lowest
 and highest environment bin where there is a small decline in the fraction of
green valley galaxies. This was also observed by \citet{Wetzel2012} for the green valley defined using  H$\alpha$ based SFRs. In Table \ref{density-sSFR-table}, we
summarise the percentage of star forming, green valley, and quenched
galaxies in low, intermediate, and high density environments where we
observe the same trend as in Figure \ref{density-sSFR}. This shows a
strong effect of environment in quenching of galaxies.

The fraction of
star-forming, green valley, and quenched galaxies for fixed
environmental bins, separately for LTS, ETS, S0s, and Es is shown in Figure \ref{density-sSFR-fraction-diff1}. We see that most of the
LTS are star-forming and most of the Es are quenched in all
environments. About 10\% of all the Es are in the green-valley in all
environments. Star-forming ETS decrease with increasing environmental
density, which will get quenched and hence the quenched fraction will
increase. Some of them may also transform into quenched S0s which is
seen as an increase in the quenched fraction of S0s. Interestingly,
S0s in lowest environmental density bin are mostly star-forming (the exact fraction however has very large uncertainity) and there are no green valley, and quenched S0s. Further, the fraction of star-forming
and green valley S0s show a gradual decline with increasing
environmental density.

Finally, we report the percentage of galaxies in low density,
intermediate density, and high density environments which are in the
star-forming, green valley, and quenched region in Table
\ref{sSFR-density-table}. As expected, the percentage of star-forming
galaxies is highest at low/intermediate densities and drastically
decreases at high densities. The percentage of green valley galaxies
increases from low to intermediate densities and then shows a
significant decrease at high densities. The percentage of quenched
galaxies increases from low to intermediate densities and then
decreases at high densities.

\subsection{Partial correlation coefficient between morphological T-type, sSFR, and $\Sigma$} \label{partial corr coeff}

In the previous sections, we studied the relation between each of the
three parameters, morphology, sSFR, and environment, taken two at a
time, and then studied the differential effect of the third
parameter. In this section, we quantify these relations using the
partial correlation coefficient (PCC).

The PCC measures the correlation between any two variables by removing
the effect of a set of control variables. For instance, in our study
we know that morphological T-type and \sSFR\ are correlated (as the
T-type increases \sSFR\ increases). However, we know that T-Type and
environmental density ($\Sigma$) is anti-correlated because of the
morphology-density relation (as $\Sigma$ increases the fraction of
early-type galaxies increases i.e. T-type decreases). And $\Sigma$ and
\sSFR\ are also anti-correlated, since as $\Sigma$ increases, \sSFR\
decreases. Thus, it is possible that the correlation between
morphology and \sSFR\ is caused by the effect of environment. In other words, because it is more likely to find early-type galaxies in high density environments, and because
high density environments are more efficient at quenching galaxies, we
find that early-type galaxies are mostly quenched. Similarly, it is
possible that we see a morphology-density relation because as the
density increases quenching becomes more efficient, and hence the
\sSFR\ decreases, and because low levels of \sSFR\ is found mainly in
early-type galaxies, we see a anti-correlation between morphology and
density . A similar case can be made of the other two variables.

Thus,
we find the PCC between: morphological T-type-\sSFR\ with $\Sigma$ as
the control variable, morphological T-type-$\Sigma$ with \sSFR\ as the
control variable, and \sSFR-$\Sigma$ with morphological T-type as the
control variable for our final sample of 7079 galaxies.

The PCC between any two random variables A and B with C as the control variable is calculated using, 
\begin{equation}
\rho_{AB\cdot C} = \frac{\rho_{AB} - \rho_{AC}\rho_{CB}}{\sqrt{1 - \rho_{AC}^2} \sqrt{1 - \rho_{CB}^2}},
\end{equation} 
where $\rho_{AB}$ is the Spearman's rank correlation coefficient between the variables A and B. 

\begin{table}
\begin{center}
\caption{Results from the partial correlation coefficient analysis. The errors on the correlation coefficients are calculated using the jacknife method.} \label{pcorr table}
\begin{tabular}{ccc}
\hline
   Variables (control variable) &  Spearman's $\rho$ & PCC \\ 
\hline
T-type-\sSFR\ ($\Sigma$) & 0.752 $\pm$ 0.003 & 0.735 $\pm$ 0.003 \\
\hline
T-type-$\Sigma$ (\sSFR) & -0.256 $\pm$ 0.010 & -0.099 $\pm$ 0.011 \\ 
\hline
$\Sigma$-\sSFR\ (T-type)& -0.256 $\pm$ 0.010 & -0.100 $\pm$ 0.010\\ 
\hline
\end{tabular} 
\end{center}
\end{table}

The results of our PCC analysis are given in Table \ref{pcorr table},
for reference we also show the Spearman's rank correlation coefficent
(Spearman's $\rho$). Since we use 6194 data points for the PCC
analysis, our results are highly significant. The Spearman's $\rho$
shows that the correlation between T-type and \sSFR\ is high, which
remains so even for the PCC. This suggests that morphology correlates
very well with \sSFR, independent of the environment. The Spearman's
$\rho$ between T-type-$\Sigma$, and $\Sigma$-\sSFR\ is weakly
anti-correlated, which considerably decreases after computing the PCC.

Thus we find that for massive galaxies, morphology is the strongest
indicator of \sSFR, and that such a relationship is real and not just
an indirect effect due to the morphology-density relation and
density-sSFR relation. This shows that, atleast for massive galaxies,
the physical processes which shape the galaxy morphology are also the
most important in deciding their star forming state.

\section{Summary \& Conclusions} \label{conclusions}

In this work, we have studied the spectral energy distributions of massive
galaxies (\stmg{10.0}) along the Hubble sequence, from Es to spirals,
using multi-wavelength SED fitting using UV-optical-mid IR data from
GALEX-SDSS-2MASS-WISE, for a sample of 6194 galaxies. For these galaxies, we are able to simultaneously constrain the
recent star formation using UV data and also take into account the dust
attenuation from warm dust using IR data. We compute the star formation rate
and the stellar mass for each  galaxy in our sample using stellar population synthesis models. Further,
due to the availability of GALEX UV data, we could define a green valley
which has less contamination from the star-forming and quenched
region. For our whole sample, the detailed visual morphologies in
terms of the Hubble T-types were taken from NA10. Also for our whole
sample we had local environmental density from
\citet{Baldry2006}. Using the morphological classifications,
specific star formation, and environmental density information, we studied the
mutual dependence of each of these parameters. Our results are
summarised as follows:
 
\begin{enumerate}
\item We find that for massive galaxies in the local Universe, LTS,
  and Es are primarily star-forming and quenched, respectively. The
  green valley is mostly populated by ETS (49.1\%) and S0s (39.9\%). A
  further split of ETS into Sa, Sab, Sb and Sbcs shows that the growth
  of the bulge plays an important role in quenching. Given that these galaxies are
  more likely to host a classical bulge which is already quenched, the
  final process of quenching likely takes place in the disk. The
  typical quenching timescale of green valley disk galaxies are known
  to be large, of the order of more than 1 Gyr \citep{Schawinski2014} and hence the disk quenching must
  be a slow process (e.g., halo-quenching, strangulation). S0s, on the
  other hand, which also host a bigger bulge have a disk that may have
  quenched more rapidly (timescales of less than 250 Myrs) than ETS and hence they are more abundant in
  the quenched region. The properties of both the ETS and S0s show that
  at least at the high mass end ($M_{\text{stellar}} > 10^{10}
  \ M_\odot$), bulge growth is important for quenching, as was also
  observed by \citet{Bluck2014}.
 
\item The fraction of galaxies of different morphological types (LTS,
  ETS, S0s and Es) as a function of \sSFR\ show very different
  trends. The distribution of LTS peaks at much lower \sSFR\ and falls
  quite rapidly for \sSFR\ $> -10$. On the other hand, the fraction of
  ETS peak near the blue end of the green valley, and falls as we
  enter the green valley. Interestingly, this fall coincides with the
  rise in the fraction of S0s, which peak near the red end of the
  green valley. This points towards the morphological transformation
  from ETS to S0s. Beyond the green valley, in the quenched region,
  the fraction of S0s and ETS declines, and the fraction of Es starts dominating. Interestingly,
  we also find a second population of S0s galaxies which are actively
  star forming. A split of our sample into low, intermediate and high
  densities shows that this relation does not change significantly
  with environment. Remarkably, even the second population of star
  forming S0s is present in all the three environmental bins. This
  supports the view that morphology is a strong indicator of the sSFR
  of a galaxy, with a weaker dependence on the environment.

\item The fraction of star forming, green valley, and quenched
  galaxies for a given morphological T-type shows that as we go from
  Es to late type galaxies, the quenched fraction decreases. As was
  earlier observed, the fraction of green valley galaxies rises from
  S0- to Sab galaxies, showing that, in the local Universe, they are
  the galaxies in transition from star forming to the quenched
  region. Remarkably, the quenched fraction of S0- and S0 galaxies is
  much higher compared to the quenched fraction of S0/a galaxies,
  which is more closer to that of Sa spirals. Furthermore, when we
  split this relation into low, intermediate and high densities, we
  find that at low densities, the fraction of star forming S0/a
  galaxies is higher than the quenched fraction. Further, at high
  densities there is a rise in the quenched fraction of S0/a
  galaxies. Also, the quenched/star-forming fraction of S0/a galaxies
  more closely follows the Sa galaxies in all environments, and unlike
  the S0- and S0 galaxies, who have a much higher quenched fraction in
  all environments. This indicates that S0/a galaxies undergo a
  different quenching process than that of S0-, and S0 galaxies.

\item We observe the morphology-density relation for our sample of
  galaxies, with the exception that the fraction of S0s remain roughly
  constant as we go to higher densities. This is likely because, we only have
  high mass S0s in our sample. The split of the morphology-density
  relation in terms of star-forming, green valley and quenched
  galaxies shows that there are green valley galaxies in all
  environments and they are mostly dominated by ETS, and
  S0s. The fraction of green valley Es rise with
  increasing environmental density, suggesting that they may have
  undergone recent gas rich major/minor merger(s). Interestingly, we
  also see the presence of star forming S0s in all environments, from
  low to high densities.

\item The quenched fraction increases with increasing environmental
  density, and except for the lowest and the highest density bin, the fraction of
  green valley galaxies remains constant. This shows that, in the
  local Universe, massive galaxies are recently undergoing transition
  in all environments. A split of this relation with different
  morphological class (LTS, ETS, S0s, and Es) shows that ETS and S0s
  are the galaxies most affected by environment. The fraction of star
  forming ETS decline with increasing density, some of which move into the
  green valley and quenched regions. This decline coincides with the
  increase in the fraction of quenched S0s, showing that possibly some
  of the quenched ETS also undergo a morphological transformation into
  S0s. Star forming S0s are most abundant in the lowest density bin,
  and they decrease as we go to higher densities.

\item Because morphology, \sSFR, and environmental density are
  correlated with each other, we find the partial correlation
  coefficient by taking two of the variables with the third variable
  as the control parameter. Our analysis shows that the correlation
  between morphology and \sSFR\ is strongest, and independent of the
  environment. Whereas the anti-correlations between
  morphology-density, and density-sSFR are both weaker when we remove the
  effect of the third parameter.
\end{enumerate}

In conclusion, we studied the star formation history using
multi-wavelength data from UV-optical-mid IR along the Hubble sequence
for a sample of 6194 galaxies. Our results shows that, in the local
Universe, for massive galaxies, ETS and S0s are currently undergoing
transition from star-forming to the quenched region.  There are two
populations of S0s, a quenched and an actively star forming one which
 is present in all environments. Further, our study of
morphology, star formation and environment shows that morphological
T-type is the strongest indicator of the sSFR of a galaxy independent
of the environment. The impact of environment in deciding whether a
galaxy is quenched or not, is seen mainly on S0/a,Sa, and Sab type
galaxies.

\section*{Acknowledgements}

We thank the anonymous referee for insightful comments that have
improved both the content and presentation of this paper. YW thanks IUCAA for hosting him on his sabbatical when the early part of this
work was completed. SB would like to acknowledge support from the
National Research Foundation research grant (PID-93727). SB, YW and OB
acknowledge support from a bilateral grant under the Indo-South Africa
Science and Technology Cooperation (PID-102296) funded by Departments
of Science and Technology (DST) of the Indian and South African
Governments.

GALEX (Galaxy Evolution Explorer) is a NASA Small Explorer, launched
in 2003 April. We gratefully acknowledge NASA's support for
construction, operation, and science analysis for the GALEX mission,
developed in cooperation with the Centre National d'Études Spatiales
of France and the Korean Ministry of Science and Technology. 

This research has made use of the NASA/IPAC Extragalactic Database
(NED), which is operated by the Jet Propulsion Laboratory, California
Institute of Technology (Caltech) under contract with NASA. 

Funding for SDSS-III has been provided by the Alfred P. Sloan
Foundation, the Participating Institutions, the National Science
Foundation, and the U.S. Department of Energy Office of Science. 
The SDSS-III web site is http://www.sdss3.org/.

SDSS-III is managed by the Astrophysical Research Consortium for the
Participating Institutions of the SDSS-III Collaboration including the
University of Arizona, the Brazilian Participation Group, Brookhaven
National Laboratory, University of Cambridge, Carnegie Mellon
University, University of Florida, the French Participation Group, the
German Participation Group, Harvard University, the Instituto de
Astrofisica de Canarias, the Michigan State/Notre Dame/JINA
Participation Group, Johns Hopkins University, Lawrence Berkeley
National Laboratory, Max Planck Institute for Astrophysics, Max Planck
Institute for Extraterrestrial Physics, New Mexico State University,
New York University, Ohio State University, Pennsylvania State
University, University of Portsmouth, Princeton University, the
Spanish Participation Group, University of Tokyo, University of Utah,
Vanderbilt University, University of Virginia, University of
Washington, and Yale University.

This publication makes use of data products from the Two Micron All
Sky Survey, which is a joint project of the University of
Massachusetts and the Infrared Processing and Analysis
Center/California Institute of Technology, funded by the National
Aeronautics and Space Administration and the National Science
Foundation. This publication makes use of data products from the
Wide-field Infrared Survey Explorer (WISE), which is a joint project
of the University of California, Los Angeles, and the Jet Propulsion
Laboratory/California Institute of Technology, funded by the National
Aeronautics and Space Administration.

\bibliographystyle{mnras}
\bibliography{references}
\end{document}